\documentclass[preprint]{emulateapj}
\usepackage{color}
\usepackage{hyperref}
\hypersetup{
    colorlinks,%
    citecolor=blue,%
    linkcolor=blue,%
    urlcolor=blue
}

\begin{document}

\title{Simulations of particle acceleration beyond the classical synchrotron burnoff limit in magnetic reconnection: An explanation of the Crab flares.}

\shorttitle{Particle acceleration beyond the synchrotron burnoff limit in magnetic reconnection.}

\author{B.~Cerutti$^{1}$, G.~R.~Werner$^{1}$, D.~A.~Uzdensky$^{1}$ \& M.~C.~Begelman$^{2,3}$} \shortauthors{Cerutti, Werner, Uzdensky, \& Begelman}

\affil{$^1$ Center for Integrated Plasma Studies, Physics Department, University of Colorado, UCB 390, Boulder, CO 80309-0390, USA; benoit.cerutti@colorado.edu, greg.werner@colorado.edu, uzdensky@colorado.edu}

\affil{$^2$ JILA, University of Colorado and National Institute of Standards and Technology, UCB 440, Boulder, CO 80309-0440, USA; mitch@jila.colorado.edu}

\affil{$^3$ Department of Astrophysical and Planetary Sciences, University of Colorado, UCB 391, Boulder, CO 80309-0391, USA}

\begin{abstract}
It is generally accepted that astrophysical sources cannot emit synchrotron radiation above $160~$MeV in their rest frame. This limit is given by the balance between the accelerating electric force and the radiation reaction force acting on the electrons. The discovery of synchrotron gamma-ray flares in the Crab Nebula, well above this limit, challenges this classical picture of particle acceleration. To overcome this limit, particles must accelerate in a region of high electric field and low magnetic field. This is possible only with a non-ideal magnetohydrodynamic process, like magnetic reconnection. We present the first numerical evidence of particle acceleration beyond the synchrotron burnoff limit, using a set of 2D particle-in-cell simulations of ultra-relativistic pair plasma reconnection. We use a new code, {\tt Zeltron}, that includes self-consistently the radiation reaction force in the equation of motion of the particles. We demonstrate that the most energetic particles move back and forth across the reconnection layer, following relativistic Speiser orbits. These particles then radiate $>160~$MeV synchrotron radiation rapidly, within a fraction of a full gyration, after they exit the layer. Our analysis shows that the high-energy synchrotron flux is highly variable in time because of the strong anisotropy and inhomogeneity of the energetic particles. We discover a robust positive correlation between the flux and the cut-off energy of the emitted radiation, mimicking the effect of relativistic Doppler amplification. A strong guide field quenches the emission of $>160~$MeV synchrotron radiation. Our results are consistent with the observed properties of the Crab flares, supporting the reconnection scenario.
\end{abstract}


\section{Introduction}\label{intro}

The maximum energy reached by a charged particle in a given astrophysical object is limited by the size of the acceleration region \citep{1984ARA&A..22..425H}. If the relativistic Larmor radius of the particle $R$ is of order the system size $L$, the particle escapes and is no longer accelerated. The maximum energy is then given by $\mathcal{E}_{\rm max}\lesssim qBL$, where $q$ is the charge of the particle and $B$ is the typical magnetic field strength. Radiative losses within the accelerator decrease this limit (see, e.g., \citealt{2002PhRvD..66b3005A,2003PhRvE..67d5401M}). The maximum energy is then set by the balance between the electric acceleration rate and the radiative power lost by the particle. In the case of synchrotron cooling, this balance leads to a remarkable result: the maximum synchrotron photon energy emitted by an electron depends only on the ratio of the electric field to magnetic field perpendicular to the particle's motion, i.e., $\epsilon^{\rm max}_{\rm sync}=(9mc^2/4\alpha_{\rm F})(E/B_{\perp})$, where $\alpha_{\rm F}$ is the fine structure constant and $mc^2$ is the rest mass energy of the electron. Hence, under ideal magnetohydrodynamic (MHD) conditions where $E\leq B$, the energy of synchrotron radiation should not exceed the fundamental constant $9mc^2/4\alpha_{\rm F}\approx 160~$MeV \citep{1983MNRAS.205..593G,1996ApJ...457..253D,2010MNRAS.405.1809L,2011ApJ...737L..40U}. An electron with energy above the radiation reaction limit would lose most of its energy in a fraction of a Larmor gyration. It is then impossible to have electrons radiating synchrotron radiation above 160~MeV with classical models of particle acceleration, all based on ideal MHD (e.g., diffuse shock acceleration), unless the plasma has a relativistic bulk motion with respect to the observer, or the electrons are the by-product of energetic particle decay.

Yet, the gamma-ray space telescopes {\em Agile} and {\em Fermi} discovered several powerful gamma-ray flares from the Crab Nebula \citep{2011Sci...331..736T,2011Sci...331..739A,2011A&A...527L...4B, 2011ApJ...741L...5S,2012ApJ...749...26B,2013ApJ...765...52S, 2013ATel.4855....1O,2013ATel.4856....1S}, presumably during which PeV electrons and positrons emit synchrotron radiation well above the $160~$MeV limit. The most powerful flare, recorded in April 2011, showed clear evidence for synchrotron emission up to $375~$MeV \citep{2012ApJ...749...26B}. This discovery suggests that an extreme and non-conventional particle acceleration mechanism is at work somewhere in the nebula, unless the emission is substantially Doppler-boosted by a factor $\gtrsim 2$. However, the typical flow velocity in the nebula (about half the speed of light, e.g., \citealt{2002ApJ...577L..49H}) and its orientation with respect to the observer give a Doppler factor of order unity. The precise location of the flare is unknown because the nebula is not resolved in gamma rays, but the lack of pulsations suggests it does not originate 
very close to the pulsar. The $\lesssim 8~$hours flux-doubling timescale observed during the flare indicates that a tiny fraction of the nebula is involved. The flaring region radiates about 30~times more than the quiescent emission above 100~MeV, which represents up to 1\% of the pulsar spin-down power \citep{2012ApJ...749...26B}. To explain particle acceleration and emission within the overall duration of the flares, ranging from a few days to a few weeks, the magnetic field should be of order a few mG, i.e., much more intense than the average $\sim 200\mu$G traditionally inferred from spectral modeling \citep{2004ESASP.552..439H,2010A&A...523A...2M}. Simultaneous observations in radio, near-infrared, optical, X-rays and in TeV gamma-rays, were not able to detect a solid counterpart to the flares (see \citealt{2013ApJ...765...56W}, and references therein), suggesting that the emitting particle spectrum is very hard, perhaps monoenergetic. It is very difficult to reconcile these puzzling features with classical models of particle acceleration and models of pulsar wind nebulae (\citealt{1974MNRAS.167....1R,1984ApJ...283..694K}, and see \citealt{2009ASSL..357..421K,2012SSRv..173..341A} for recent reviews).

Various models have been proposed to solve the Crab flares mystery. Several studies invoke a relativistic Doppler boosting of the flaring region by a factor of a few. It was proposed that a mildly relativistic flow could be achieved close to the pulsar wind termination shock \citep{2011MNRAS.414.2017K,2012MNRAS.422.3118L,2011MNRAS.414.2229B}, in relativistic reconnection events within the nebula \citep{2012MNRAS.426.1374C}, in a magnetized flow at the base of the Crab jets \citep{2012MNRAS.427.1497L}, or in knots of energetic particles \citep{2011ApJ...730L..15Y}. The dissipation of the striped pulsar wind structure through the shock \citep{2007A&A...473..683P,2011ApJ...741...39S} could also generate rapidly fluctuating magnetic field on small scales resulting in synchrotron gamma-ray flares \citep{2012MNRAS.421L..67B}. If the magnetic turbulence in the nebula occurs on a length scale shorter than the synchrotron photon formation length $mc^2/eB$, then the particles could emit jitter radiation \citep{2000ApJ...540..704M}, with typical energy greater than the classical synchrotron limit \citep{2013ApJ...763..131T}. Alternatively, particle acceleration could occur in regions of strong coherent electric field, in twisted toroidal fields \citep{2012ApJ...751L..32S}, or in magnetic reconnection sites within the nebula \citep{2011ApJ...737L..40U,2012ApJ...746..148C}.

\begin{figure}
\epsscale{1.2}
\plotone{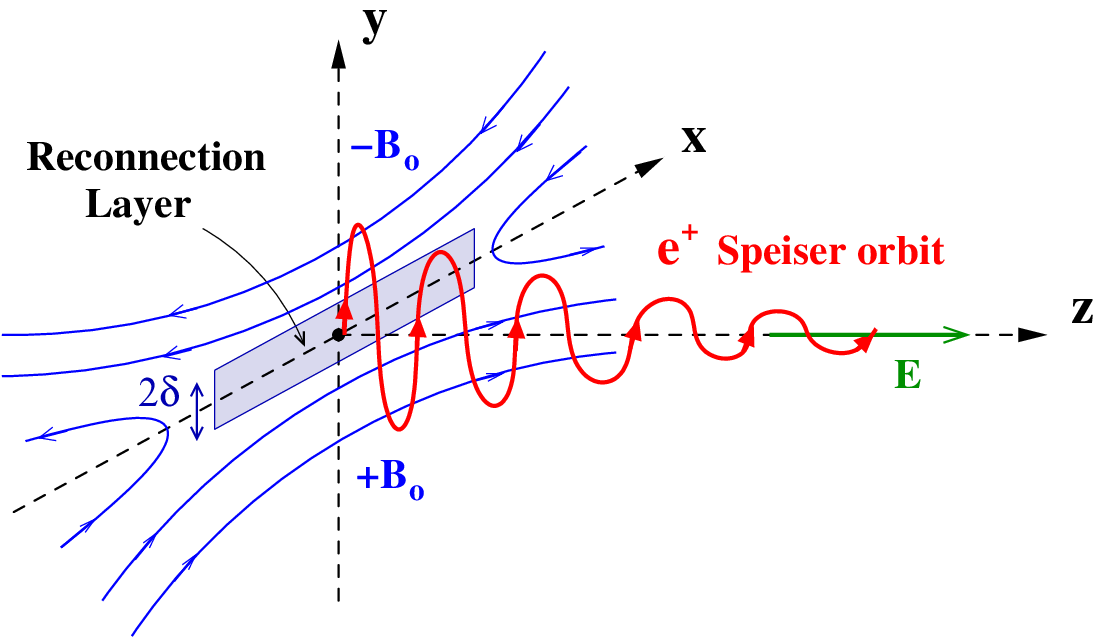}
\caption{This diagram represents a relativistic Speiser orbit, i.e., the trajectory of a charged particle (here a positron) moving back and forth across the reconnection layer of some thickness $2\delta$. The particle is accelerated along the $z$-direction by the reconnection electric field, $E$. The initial reconnecting magnetic field is along the $\pm x$-directions ($\pm B_0$), and reverses across the $y=0$ plane.}
\label{fig_intro}
\end{figure}

Magnetic reconnection offers natural locations (within the diffusion region where the magnetic field is small and reverses) in which the electric field can exceed the magnetic field. In principle, it is possible to accelerate particles above the classical radiation reaction limit at these sites \citep{2004PhRvL..92r1101K}. \citet{2011ApJ...737L..40U} demonstrated that the highest energy particles are trapped and focused towards the reconnection layer midplane\footnote{A similar phenomenon occurs in man-made accelerators, where gradients of magnetic fields are generated to confine and focus particle orbits, see e.g., \citet{1958AnPhy...3....1C}.} (see also \citealt{2007A&A...472..219C}), following the relativistic analog of Speiser orbits (\citealt{1965JGR....70.4219S}, see Fig.~\ref{fig_intro}). Once deep inside the layer, the particles are subject to weak radiative losses, but strong coherent electric field. The layer acts as a linear accelerator, and the maximum energy of the particles is then limited just by the total electric potential drop along the layer, i.e., $\mathcal{E}_{\rm max}\sim eEL$, where $L$ is the length of the layer. Using relativistic test-particle simulations, we found in \citet{2012ApJ...746..148C} that magnetic reconnection could generate a quasi-monoenergetic beam of particles above the radiation reaction limit in the Crab Nebula.

In this paper, we re-examine particle acceleration in ultra-relativistic pair plasma reconnection, using 2D particle-in-cell (PIC) simulations. Such simulations capture self-consistently the time-evolution of fields and particles at the kinetic level. For this study, we developed a new relativistic PIC code, called {\tt Zeltron}, that includes the radiation reaction force on the particles. Our first objective is to investigate ab initio whether reconnection can accelerate particles above the radiation reaction limit, under realistic physical conditions. Our second objective is to study the radiative signature of such acceleration following \citet{2012ApJ...754L..33C}, in order to explain all observed features of the Crab flares. \citet{2012ApJ...754L..33C} included optically thin synchrotron radiation as a tracer, but this calculation was not self-consistent because it did not treat radiation reaction effects on particle motion. Section~\ref{zeltron} describes the main capabilities of {\tt Zeltron}, with an emphasis on the radiation reaction force. Section~\ref{init} gives the initial setup of the simulations performed in this study, and Section~\ref{sim} presents the main results on particle acceleration and radiation. We discuss our findings in the context of the Crab gamma-ray flares in Section~\ref{crab}. Section~\ref{ccl} summarizes the main results of this work.

\section{The particle-in-cell code {\tt Zeltron}}\label{zeltron}

{\tt Zeltron} is a new three-dimensional, parallel (domain decomposition with MPI), relativistic, electromagnetic particle-in-cell code developed independently from other existing codes (for a review about PIC methods, see e.g., \citealt{2003LNP...615....1P,Birdsall...Langdon}). {\tt Zeltron} follows an explicit finite-difference scheme on a Cartesian grid, with a time step $\Delta t$ set at a fraction of the maximum stable step determined by the Courant-Friedrichs-Lewy condition, i.e., $c\Delta t\leq c\Delta t_{\rm CFL}=(1/\Delta x^2+1/\Delta y^2+1/\Delta z^2)^{-1/2}$, where $c$ is the speed of light, $\Delta x,~\Delta y,~$and $\Delta z$ are the minimum grid spacing in the $x$-, $y$-, and $z$-directions. The code uses the Yee algorithm \citep{1966ITAP...14..302Y} to solve the time-dependent Maxwell's equations, given by
\begin{eqnarray}
\frac{\partial\mathbf{E}}{\partial t} &=& c\mathbf{\nabla}\times\mathbf{B}-4\pi\mathbf{J}\\
\frac{\partial\mathbf{B}}{\partial t} &=& -c\mathbf{\nabla}\times\mathbf{E},
\end{eqnarray}
where $\mathbf{E}$ is the electric field, $\mathbf{B}$ is the magnetic field, and $\mathbf{J}$ is the current density. This algorithm has second-order error in space and time and ensures that $\mathbf{\nabla}\cdot\mathbf{B}=0$ at any instant of the simulation (to the computer round-off accuracy). The numerical scheme does not, however, satisfy the Maxwell-Gauss equation $\mathbf{\nabla}\cdot\mathbf{E}=4\pi\rho$ exactly, where $\rho$ is the charge density. The electric field has to be corrected every time step by the small amount $\delta\mathbf{E}$, obtained by solving Poisson's equation \citep{2003LNP...615....1P}, i.e.,
\begin{equation}
\nabla^2\left(\delta\phi\right)=-\left(4\pi\rho-\mathbf{\nabla}\cdot\mathbf{E}\right),
\end{equation}
and $\delta\mathbf{E}=-\mathbf{\nabla}\delta\phi$. The Poisson solver implemented in the code utilizes an iterative Gauss-Seidel method (with 5-points in 2D, and 7-points in 3D).

The main novelty of {\tt Zeltron}, compared to most PIC codes\footnote{Several PIC codes used in the laser-plasma interaction community (see e.g., \citealt{2002PhRvL..88r5002Z,2009PhPl...16i3115S,2010NJPh...12l3005T, 2012PhRvE..86c6401C}) do include the radiation reaction force, in preparation for future radiation pressure-dominated plasma experiments with the new generation of ultra-intense lasers. We note also that \citet{2009PhRvL.103g5002J} present the first study of radiation-dominated reconnection in the relativistic regime, using PIC simulations with radiation reaction force.}, is the ability to take into account the effect of the radiation reaction force on the motion of the particles. The (non-covariant) equation of motion of a single particle is given by the Lorentz-Abraham-Dirac equation \citep{1975ctf..book.....L}
\begin{equation}
mc\frac{d\mathbf{u}}{dt}=q\left(\mathbf{E_{\rm i}}+\frac{\mathbf{u}\times\mathbf{B_{\rm i}}}{\gamma}\right)+\mathbf{g},
\end{equation}
where $\mathbf{u}=\gamma\mathbf{v}/c$ is the four-velocity of the particle, $\gamma=\left(1-v^2/c^2\right)^{-1/2}$ is the associated Lorentz factor, and $\mathbf{g}$ is the radiation reaction force. The fields at the location of the particle $\mathbf{E_{\rm i}}$ and $\mathbf{B_{\rm i}}$, are linear interpolations of the fields $\mathbf{E}$ and $\mathbf{B}$ known at the grid nodes. Within the framework of classical electrodynamics, the radiation reaction force is obtained from the Landau-Lifshitz equation \citep{1975ctf..book.....L}, valid as long as the product $\gamma B$ is much smaller than the quantum critical magnetic field $B_{\rm QED}=m^2c^3/e\hbar=4.4\times 10^{13}~$G, where $e$ is the elementary electric charge, and $\hbar=h/2\pi$ with $h$ the Planck constant. In the ultra-relativistic regime ($\gamma\gg 1$), the radiation reaction force can simply be expressed as a continuous friction force opposite to the particle's direction of motion (see discussion in \citealt{
2012ApJ...746..148C}), i.e.,
\begin{equation}
\mathbf{g} = -\frac{2}{3}r_{\rm e}^2\gamma\left[\left(\mathbf{E_{\rm i}}+\frac{\mathbf{u}\times\mathbf{B_{\rm i}}}{\gamma}\right)^2-\left(\frac{\mathbf{u}\cdot\mathbf{E_{\rm i}}}{\gamma}\right)^2\right]\mathbf{u},
\label{rad_force}
\end{equation}
where $r_{\rm e}=e^2/mc^2$ is the classical radius of the electron. Following \citet{2010NJPh...12l3005T}, {\tt Zeltron} uses a modified Boris algorithm \citep{Birdsall...Langdon} to solve the equation of motion with the radiation reaction force given in Eq.~(\ref{rad_force}). For an alternative implementation of the radiation reaction force in PIC codes, see also \citet{2009PhPl...16i3115S,2012PhRvE..86c6401C}. 

The code uses linear interpolation to deposit the charges and currents generated by each particle at the nodes of the computational grid, and computes the charge and current densities for Maxwell's equations. The code assigns variable weights to the macro-particles to model particle density gradients. {\tt Zeltron} does not strictly conserve the total energy. For the purpose of this study, we ran {\tt Zeltron} successfully on thousands of cores on the Kraken supercomputer\footnote{National Institute for Computational Sciences (\url{www.nics.tennessee.edu/}).} and on the University of Colorado Janus and Verus supercomputers. We find excellent agreement between {\tt Zeltron} and the well-tested PIC code {\tt Vorpal} \citep{2004JCoPh.196..448N}, in the limit where radiative losses are neglected (the radiation reaction force is currently not implemented in {\tt Vorpal}).

\section{Setup of the simulations}\label{init}

\begin{figure*}
\epsscale{1.0}
\plotone{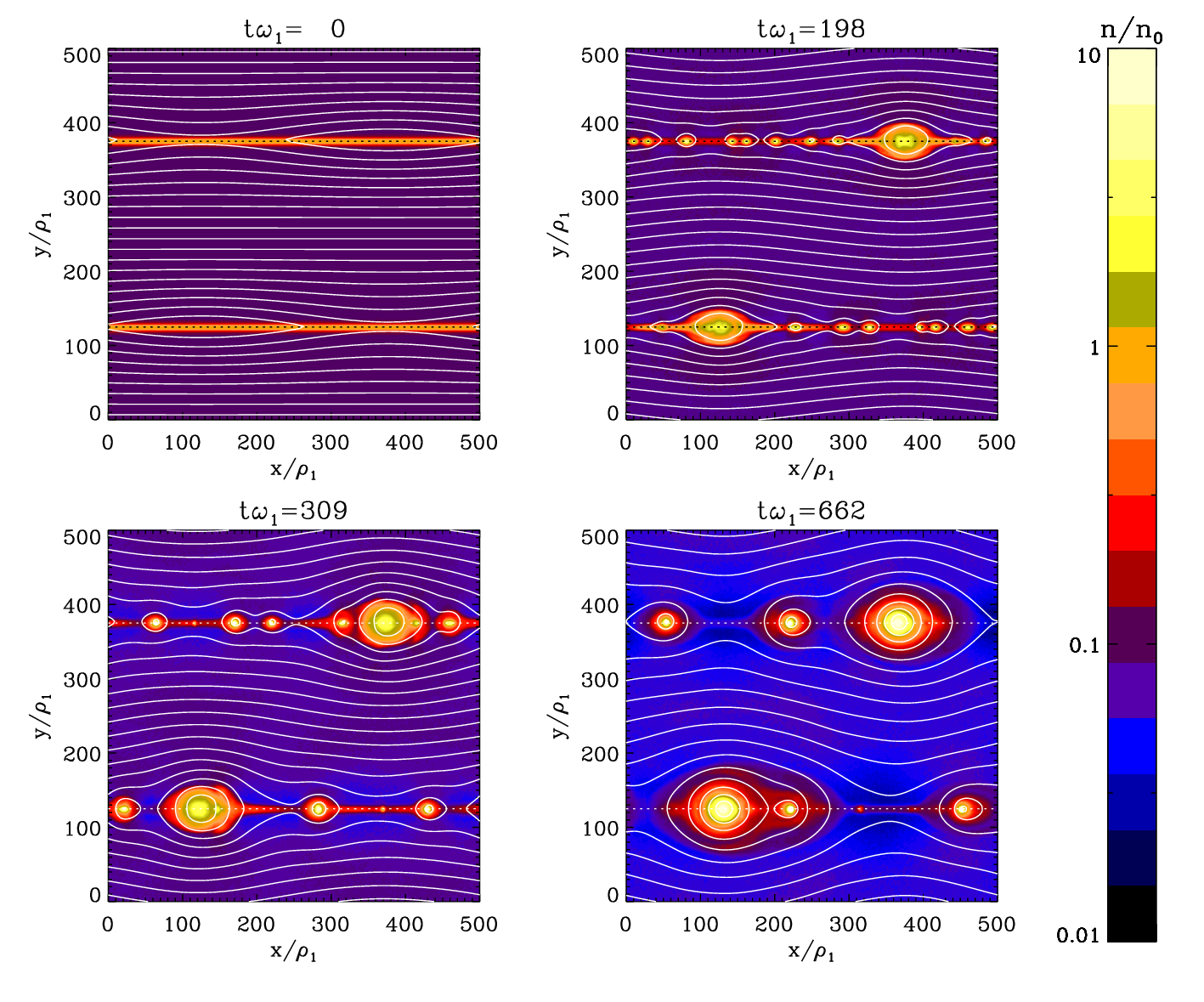}
\caption{Plasma density (color-coded) normalized to the initial drifting particle density $n_0$, and magnetic flux function iso-contours tracing magnetic field lines (white solid lines), shown at $t\omega_1=0$ (top-left), $t\omega_1=198$ (top-right), $t\omega_1=309$ (bottom-left), and $t\omega_1=662$ (bottom-right).}
\label{fig_density}
\end{figure*}

We present a series of simulations of ultra-relativistic electron-positron pair plasma reconnection with radiation reaction force in 2.5D (the fields depend on 2 coordinates, but the motion of particles is 3D), using {\tt Zeltron}. The initial setup is very similar to our previous study \citep{2012ApJ...754L..33C}, which is standard in such reconnection simulations \citep{2001ApJ...562L..63Z,2007ApJ...670..702Z,2008ApJ...677..530Z, 2004PhPl...11.1151J,2007PhPl...14e6503B,2012ApJ...750..129B, 2007PhPl...14g2303D,2007A&A...473..683P,2008ApJ...682.1436L, 2008PhRvL.101l5001Y,2009PhRvL.103g5002J,2011PhPl...18e2105L,
2011ApJ...741...39S, 2012arXiv1208.0849K}. The computational domain is a rectangle of size $L_{\rm x}\times L_{\rm y}$, where $L_{\rm x,y}$ is the length of the box in each direction. It contains two anti-parallel relativistic, flat, Harris current sheets \citep{2003ApJ...591..366K} in the $xz$-plane, which enables us to set periodic boundary conditions in the $x$- and $y$- directions. The reconnecting magnetic field, $B_{\rm x}$, reverses across each layer along the $y$-direction, and is given by
\begin{equation}
B_{\rm x} = \left\{ \begin{array}{lcl} -B_0\tanh\left(\frac{y-L_{\rm y}/4}{\delta}\right) &\mbox{if} & y<L_{\rm y}/2\\
~~B_0\tanh\left(\frac{y-3L_{\rm y}/4}{\delta}\right) & \mbox{if} & y>L_{\rm y}/2 \end{array} \right. ,
\end{equation}
where $B_0$ is the initial upstream reconnecting field strength, and $\delta$ is the initial layer half-thickness (Fig.~\ref{fig_intro}). The strength of the uniform guide field component ($B_{\rm z}$) is a free parameter that varies from $0$ to $B_0$ in our simulations (see Section~\ref{guide}). There is no electric field initially, $\mathbf{E}=\mathbf{0}$. We apply an initial perturbation to the magnetic field ($10\%$ maximum amplitude in the magnetic flux function, see Fig.~\ref{fig_density}), in order to force the onset of reconnection at the beginning of the simulation, thereby reducing the computing time. If there is no perturbation, we find that reconnection eventually happens spontaneously in the simulation because of the high numerical noise inherent to PIC codes.

A population of relativistic thermal pairs, following a Maxwell-J\"uttner distribution and concentrated in the current layers, balances the upstream magnetic pressure and carries the initial current in the $\pm z$-directions given by the bulk motion of these particles, drifting at a velocity $v_{\rm drift}/c=\beta_{\rm drift}=0.6$. The density profile of the drifting particles is
\begin{equation}
n_{\rm drift} = \left\{ \begin{array}{lcl} n_0\left[\cosh\left(\frac{y-L_{\rm y}/4}{\delta}\right)\right]^{-2} &\mbox{if} & y<L_{\rm y}/2\\
n_0\left[\cosh\left(\frac{y-3L_{\rm y}/4}{\delta}\right)\right]^{-2} & \mbox{if} & y>L_{\rm y}/2 \end{array} \right. ,
\label{density_drift}
\end{equation}
with
\begin{equation}
n_0 = \frac{k T_{\rm drift}\left(1-\beta_{\rm drift}^2\right)^{1/2}}{4\pi e^2 \beta_{\rm drift}^2\delta^2},
\end{equation}
where $k$ is the Boltzmann constant, and $T_{\rm drift}$ is the temperature of the drifting particles defined in their co-moving frame. In the simulation, the drifting particles are injected uniformly throughout the box, but with variable weight according to Eq.~(\ref{density_drift}). The purpose of this procedure is to represent low density distributions with less noise. In addition, we fill the box with a uniform density $n_{\rm bg}=0.1 n_0$ of isotropic non-thermal ultra-relativistic particles (see explanation below), such that
\begin{equation}
\frac{d{\rm N}_{\rm bg}}{d\gamma} = \left\{ \begin{array}{lcl} K\gamma^{-p} &\mbox{if} & \gamma_1<\gamma<\gamma_2\\
0 & \mbox{otherwise} &  \end{array} \right. ,
\end{equation}
$p$ is the index of the power-law, and $K$ is a normalization constant.

We initialize the simulations with physical parameters consistent with the conditions thought to exist in the flaring region of the Crab Nebula, although our results are general and scalable. The energy and spatial scales of the problem are obtained from the reconnecting field strength $B_0$, of order a few milliGauss during Crab flares. Following \citet{2012ApJ...746..148C}, we set $B_0=5~$mG. The fiducial radiation-reaction-limited energy of the particles, given by the balance of the electric force with the radiation reaction force for $E=B_0$, is then equal to
\begin{equation}
\gamma_{\rm rad}=\left(\frac{3 e}{2 r_{\rm e}^2 B_0}\right)^{1/2}\approx 1.3\times 10^9 B_{\rm 5mG}^{-1/2},
\label{grad}
\end{equation}
where $B_{\rm 5mG}=B_0/5~$mG, allowing electrons with energies up to in the PeV range. The Larmor radius associated with such a particle is $R_{\rm rad}=\gamma_{\rm rad}mc^2/eB_0\approx 4.6\times 10^{14}~$cm, or $4.25$~light-hours. The maximum energy expected to be reached by the particles in the simulation (in the absence of radiative losses) is limited by the electric potential drop along the layer, which is proportional to the system size $L$ and the dimensionless reconnection rate $\beta_{\rm rec}<1$, such that $\gamma_{\rm max}\lesssim \beta_{\rm rec}e B_0 L/mc^2$. To observe particle acceleration above the radiation reaction limit (Eq.~\ref{grad}), the system size $L$ must be bigger than $R_{\rm rad}$. $R_{\rm rad}$ must also be large compared with the smallest spatial scales in the system, given by the Larmor radius of the lowest energy particles injected initially. We set the minimum Lorentz factor of the background particles to $\gamma_1=4\times10^7$ with a power-law index of $p=2$ extending up to $\gamma_2=4\times 10^8<\gamma_{\rm rad}$, and an ultra-relativistic temperature $kT_{\rm drift}/mc^2=4\times 10^7$ for the drifting particles. In reality, the typical energy of the background particles in the Crab Nebula is probably much lower, of order $\gamma_1\sim 10^4$--$10^6$. However, such a broad range in energy cannot be reached with the current computing power available, even for 2D simulations. Our simulations model only the high-energy end of the Crab Nebula spectrum. The initial $-2$ power-law of the electrons approximately represents the purely non-thermal background particles emitting the quiescent emission of the Crab Nebula, inferred from spectral modeling \citep{2004ESASP.552..439H,2010A&A...523A...2M}. This assumes pre-acceleration of the particles in the nebula, possibly by shock acceleration or by other reconnection events.

In the following, we express all spatial scales in terms of the initial minimum Larmor radius $\rho_1=\gamma_1 mc^2/eB_0\approx 1.4\times 10^{13}~$cm, and timescales with respect to the corresponding inverse Larmor frequency $\omega_1^{-1}=\rho_1/c\approx 455~$seconds. We perform all the simulations with $L_{\rm x}=L_{\rm y}=500\rho_1$, which corresponds to $7\times 10^{15}$~cm or $2.7$~light-days. For a typical value of $\beta_{\rm rec}=0.1$--$0.2$, we have $\gamma_{\rm max}\lesssim 50$--$100\gamma_1\approx 2$--$4\times 10^9$, allowing us to see particles with gamma above $\gamma_{\rm rad}\approx 1.3\times 10^9$. The other relevant quantities of the plasmas are the electron skin-depth in the layer $d_{\rm e}=(kT_{\rm drift}/4\pi n_0 e^2)^{1/2}\approx 1.8\rho_1$, the initial layer thickness $\delta=2kT_{\rm drift}(1-\beta_{\rm drift}^2)^{1/2}/\beta_{\rm drift}eB_0\approx 2.7\rho_1$, and the upstream magnetization parameter $\sigma=B_0^2/4\pi n_{\rm bg} \gamma_1 mc^2\approx 16$. The computational grid is Cartesian and uniform, composed of $1440^2$ cells, giving a spatial resolution of $\approx 3$ cells per $\rho_1$ or $\approx 8$ cells per $\delta$. We note that the simulation becomes unstable at late times if $\Delta t=\Delta t_{\rm CFL}$, only when radiative cooling is strong and if there is no guide field. In particular, the electric field oscillates between adjacent time steps leads to an artificial increase of radiative losses. We find that setting the time step to $\Delta t=0.3\Delta t_{\rm CFL}$ is good enough to quench the development of this numerical instability. The time resolution is then $\Delta t\approx 0.07\omega_1^{-1}$. The total initial number of particles per cell is $100$ (25 electrons and 25 positrons for each population). Table~\ref{tab_params} gives all the physical and numerical parameters and their values chosen for this study.

\begin{table}[htp]
\caption{Physical and numerical parameters common to all the simulations reported in this work.}
\label{tab_params}
\centering
\begin{tabular}{|c|c|}
\hline
 Physical parameters & Set values\\
\hline
$B_0$ & $5~$mG \\
$\gamma_{\rm rad}$ & $1.3\times 10^9$\\
$\rho_1$ & $1.4\times10^{13}$~cm\\
$d_{\rm e}/\rho_1$ & $1.8$\\
$\delta/\rho_1$ & $2.7$\\
$\omega_1^{-1}$ & $455$~s\\
$kT_{\rm drift}/mc^2$ & $4\times10^7$\\
$\beta_{\rm drift}$ & $0.6$\\
$n_{\rm bg}/n_0$ & $0.1$\\
$\sigma$ & $16$\\
$\gamma_1$ & $4\times 10^7$\\
$\gamma_2$ & $4\times 10^8$\\
$p$ & $2$\\
\hline
Numerical parameters & Set values\\
\hline
$\rho_1/\Delta x$ & $3$\\
$\rho_1/\Delta y$ & $3$\\
$\Delta t\times\omega_1$ & $0.07$\\
Particles/cell & $100$\\
$L_{\rm x}/\rho_1$ & $500$\\
$L_{\rm y}/\rho_1$ & $500$\\
\hline
\end{tabular}
\end{table}

\section{Results of the simulations}\label{sim}

This section presents the results based on a set of simulations of size $(L_{\rm x}\times L_{\rm y})=(500\rho_1)^2$ to investigate the general properties of reconnection under strong synchrotron cooling conditions and particle acceleration above the radiation reaction limit (Section~\ref{evol}-\ref{speiser}); we compare with an identical simulation without the radiation reaction force. In Section~\ref{var}, we investigate the variability patterns (spectra, light curves, power-spectra) of the high-energy radiation emitted by the layer, and we show that these features are robust and reproducible, using $10$ identical simulations with the radiation reaction force. The statistical variations in this sample of simulations originate only from the initial random positions and velocities of the particles in the box. We present also a set of $4$ simulations to study the effect of the guide field on particle acceleration above the radiation reaction limit (Section~\ref{guide}).

\subsection{Time evolution of reconnection}\label{evol}

\begin{figure*}
\epsscale{1.0}
\plotone{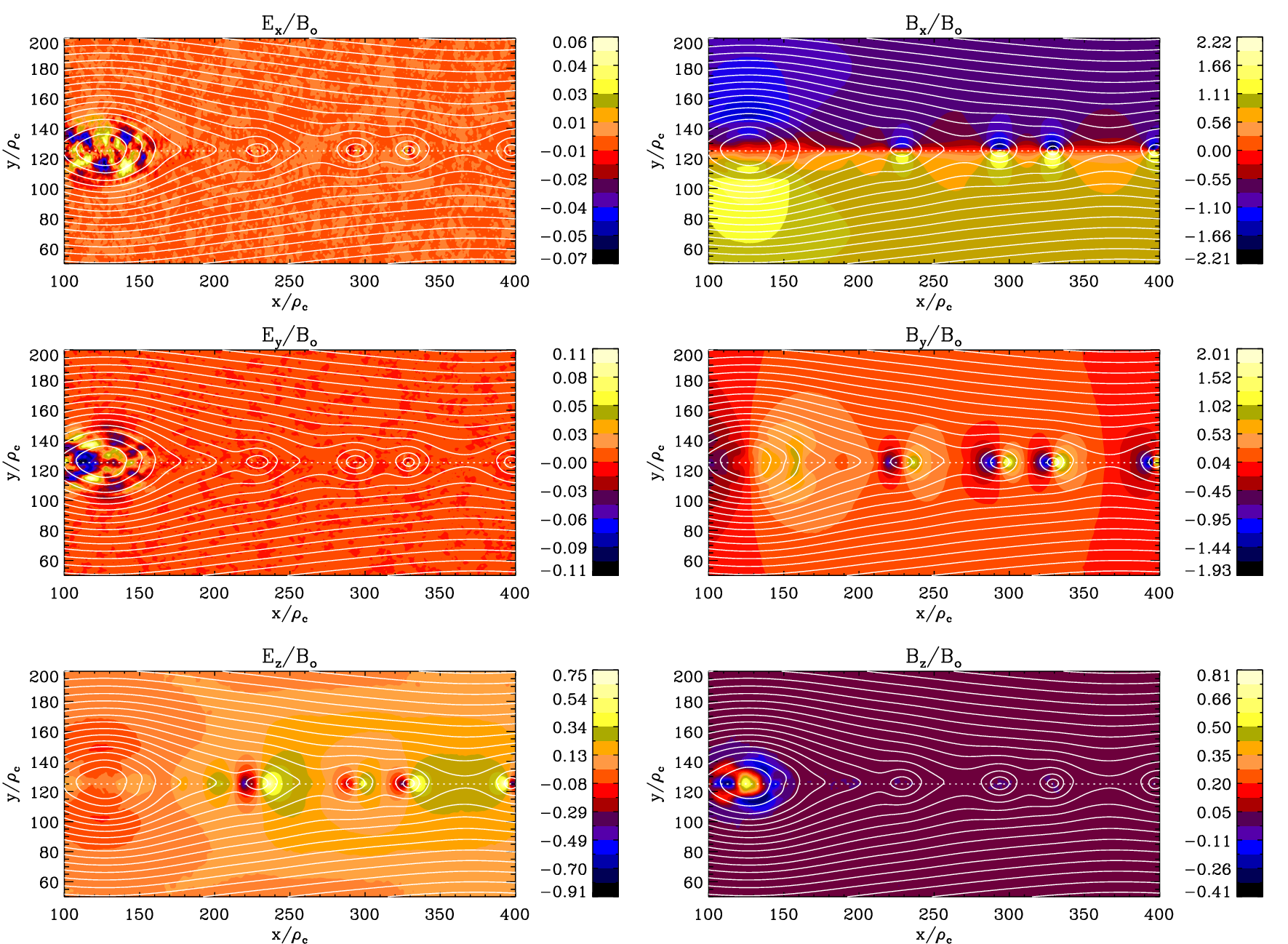}
\caption{Electric (left) and magnetic (right) field components normalized to $B_0$. The $x$- ($E_{\rm x},$ $B_{\rm x}$, top panels), $y$- ($E_{\rm y},$ $B_{\rm y}$, middle panels), and $z$- ($E_{\rm z},$ $B_{\rm z}$, bottom panels) components are shown at $t\omega_1=198$, zoomed in on the lower layer islands. White solid lines represent magnetic field lines.}
\label{fig_fields}
\end{figure*}

At the beginning of the simulation, the system remains quasi-static\footnote{This phase is significantly shortened by the initial perturbation applied to the magnetic field.} for about $t\omega_1\lesssim 110$, before the layers become unstable to tearing modes and break up into several islands of closed magnetic field loops filled with plasma (``magnetic islands'' or ``plasmoids'', see Fig.~\ref{fig_density}). This instability produces multiple secondary X-points between the plasmoids, where the magnetic field reconnects and where the electric field, mostly along the $\pm z$-directions $(E\approx E_{\rm z})$ is intense, of order $B_0$, leading to strong particle acceleration (see Fig.~\ref{fig_fields}, bottom-left panel). In addition, the formation of islands induces a strong bipolar reconnected magnetic field $B_{\rm y}\sim B_0$ concentrated at the edges of islands between two X-points (see Fig.~\ref{fig_fields}, middle-right panel). The magnetic tension of this field drives the reconnection outflow by pushing the plasma away from X-points into the direction of magnetic islands. Later on, magnetic islands become rounder and merge with each other to form bigger islands, until there is only a single big island and a single X-point left (per layer) at the end of reconnection (at $t\omega_1\gtrsim 450$, Fig.~\ref{fig_density}; the end of the simulation does not reach the full saturated state). This peculiar symmetric final state arises from the choice of double-periodic boundary conditions.

\begin{figure}
\epsscale{1.2}
\plotone{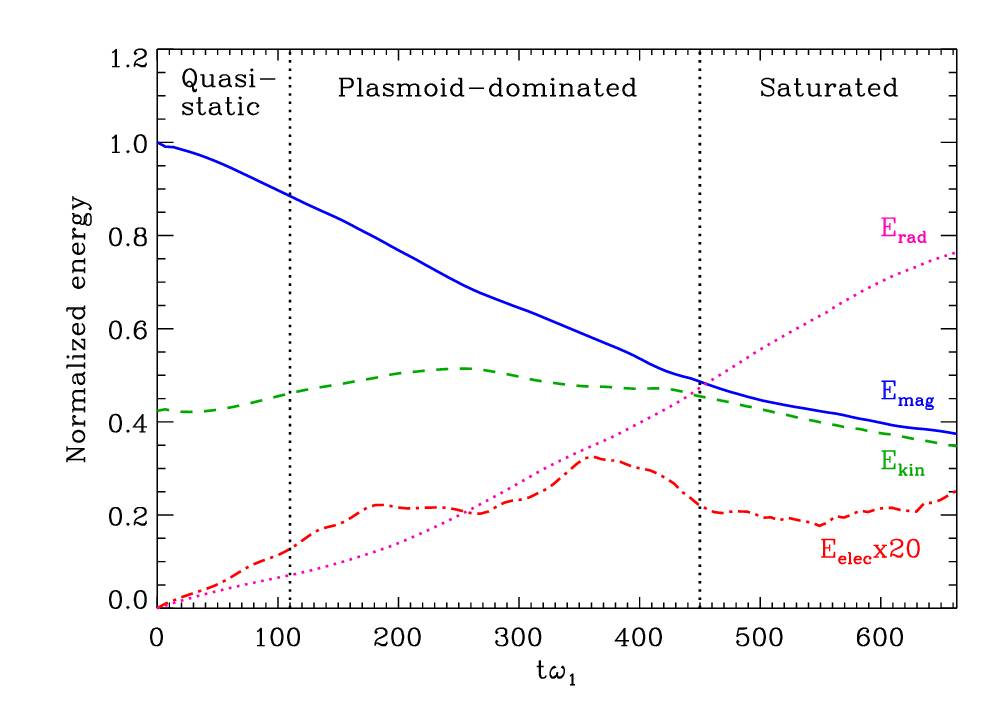}
\caption{Time evolution of the total magnetic (``$E_{\rm mag}$'', blue solid line) and electric field energies (``$E_{\rm elec}$'', multiplied by a factor $20$ for readability, red dot-dashed line), the kinetic energy of all the particles (``$E_{\rm kin}$'', green dashed line), and the total energy lost through synchrotron cooling by the particles (``$E_{\rm rad}$'', purple dotted line). The three main phases of reconnection (1. ``Quasi-static'', 2. ``Plasmoid-dominated'', and 3. ``Saturated'') are roughly delimited by vertical dotted lines.}
\label{fig_energy}
\end{figure}

We find that the reconnection rate increases by a factor of $2$ when the radiation reaction is self-consistently included, $\beta_{\rm rec}\approx 0.3$. Because the initial pressure balance is approximately maintained during the simulation, the layer compresses due to high synchrotron energy losses inside the layer, leading to an increase in the reconnection rate \citep{2011PhPl...18d2105U}. The magnetic islands tend to be smaller and denser with radiative losses. Apart from this, we do not find any qualitative difference in the overall time-evolution of the reconnection process, with or without the radiation reaction force. Fig.~\ref{fig_energy} shows the time evolution of the different energy components in the system, i.e., the energy of the fields, of the particles and of the radiation. About $63\%$ of the initial magnetic energy is dissipated and entirely radiated away by the particles at the end of the simulation. The total energy is conserved to within about $5\%$ throughout the simulation. This moderate error comes from the overestimation of the synchrotron energy losses due to the high-frequency fluctuating electric field described in Section~\ref{init}. If the radiation reaction force is neglected, or if there is a non-zero guide field, the total energy is very well conserved with less than $0.2\%$ error.

To summarize, the overall time evolution of reconnection can be divided schematically into three main phases shown in Fig.~\ref{fig_energy}:
\begin{itemize}
\item{\bf Phase~1.} Quasi-static state where the electric field builds up smoothly with little magnetic reconnection. About 12\% of the total magnetic energy is dissipated during this period.
\item{\bf Phase~2.} Plasmoid-dominated reconnection. This is the most active and bursty period of magnetic dissipation, during which about 40\% of the total magnetic energy is dissipated. This phase is characterized by the strong competition between particle acceleration and cooling.
\item{\bf Phase~3.} Saturated state, where particle cooling dominates. The particles and the magnetic field are approximately in equipartition. The reconnection rate decreases so that only 11\% of the total magnetic energy is dissipated during this period.
\end{itemize}

\subsection{Particle and synchrotron radiation spectra and anisotropies}\label{spec}

\begin{figure}
\epsscale{1.1}
\plotone{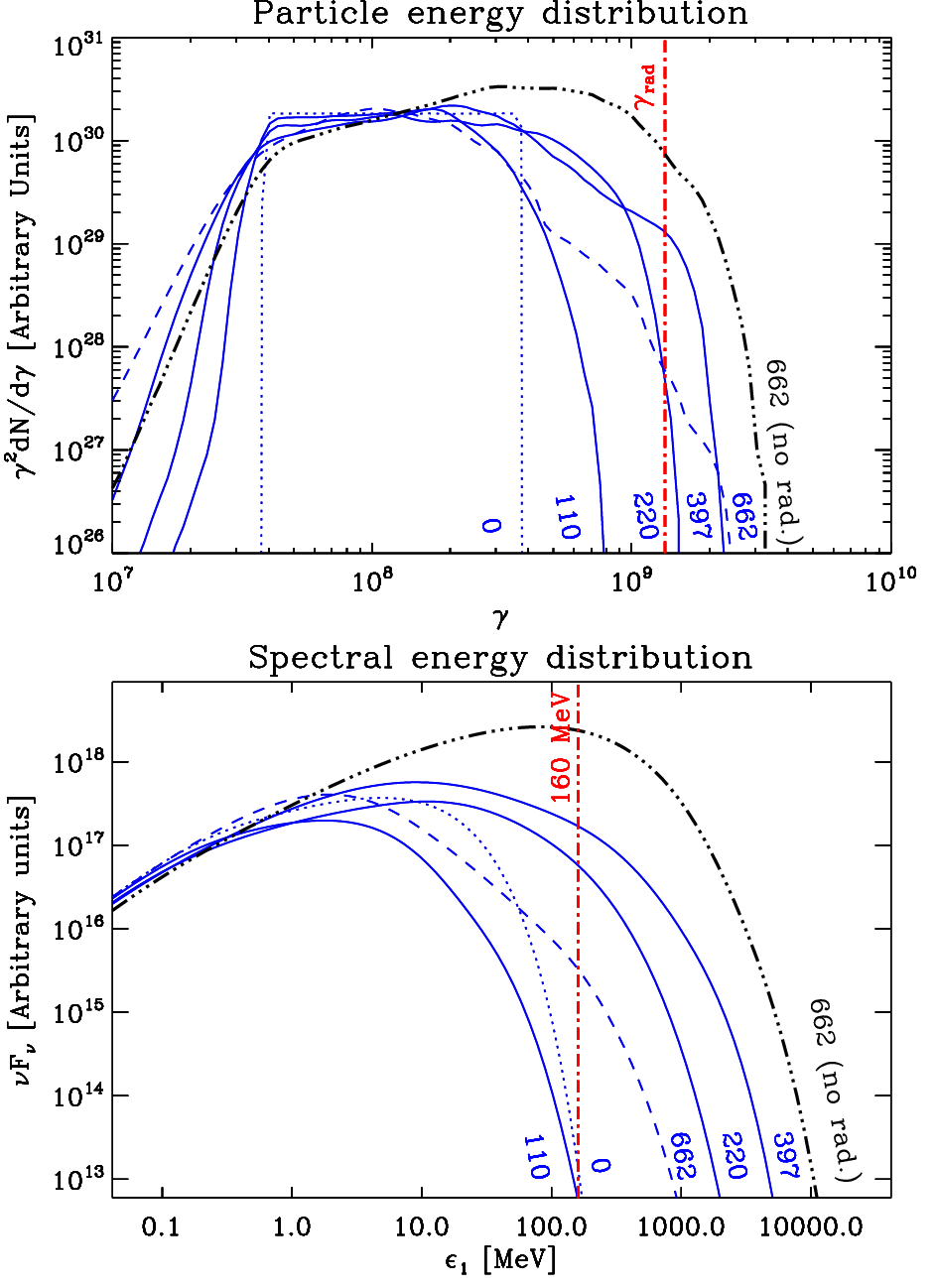}
\caption{Particle ($\gamma^2 d{\rm N}/d\gamma$, top) and photon spectral ($\nu F_{\nu}$, bottom) energy distributions averaged over all directions, at $t\omega_1=0$ (dotted line),$~110$,$~220$,$~397$, and $662$ (dashed line). The red vertical dot-dashed line in the top panel shows the classical radiation-reaction-limited energy $\gamma_{\rm rad}$ defined in Eq.~(\ref{grad}), and the corresponding $160~$MeV synchrotron photon energy limit in the bottom panel. The thick black 3-dot-dashed line marked ``no rad.'' in both panels shows the final quasi-steady energy distributions at $t\omega_1=662$, if there are no radiative losses in the simulation.}
\label{fig_spectra_iso}
\end{figure}

\begin{figure}
\epsscale{1.1}
\plotone{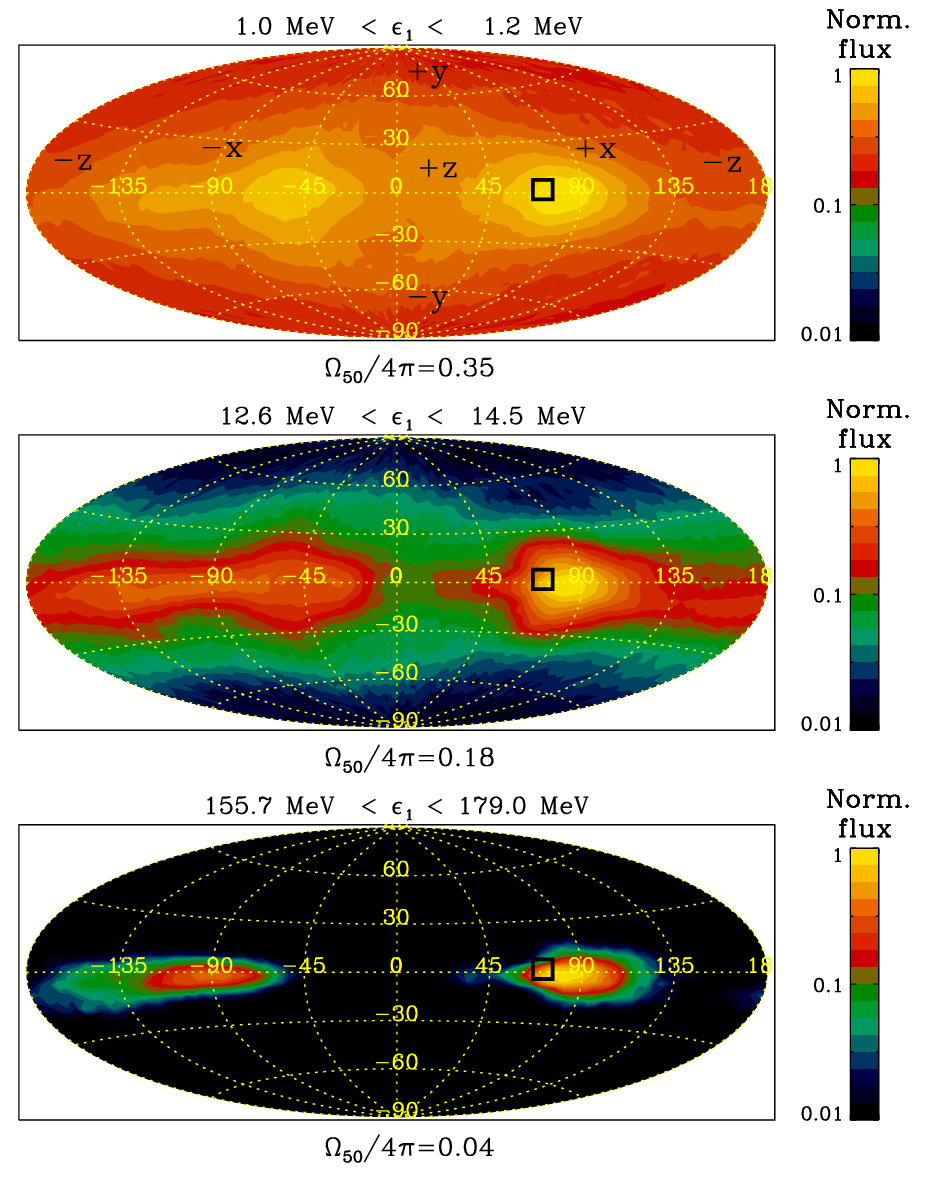}
\caption{Energy-resolved angular distribution of the synchrotron radiation flux $d(\nu F_{\nu})/d\Omega/d\epsilon_1$ emitted at $t\omega_1=397$, using the Aitoff projection. Each panel shows the angular distribution of radiation in a different photon energy band: $1~\rm{MeV}<\epsilon_1<1.2~\rm{MeV}$ (top), $12.6~\rm{MeV}<\epsilon_1<14.5~\rm{MeV}$ (middle), and $155.7~\rm{MeV}<\epsilon_1<179.0~\rm{MeV}$ (bottom). Fluxes are normalized to the maximum value in each band. The solid angle covered by half of the flux and normalized by $4\pi$, $\Omega_{50}/4\pi$, is given below each panel. The black square box indicates the direction where the anisotropic spectra are shown in Fig.~\ref{fig_spectra_anis}.}
\label{fig_angular}
\end{figure}

\begin{figure}
\epsscale{1.1}
\plotone{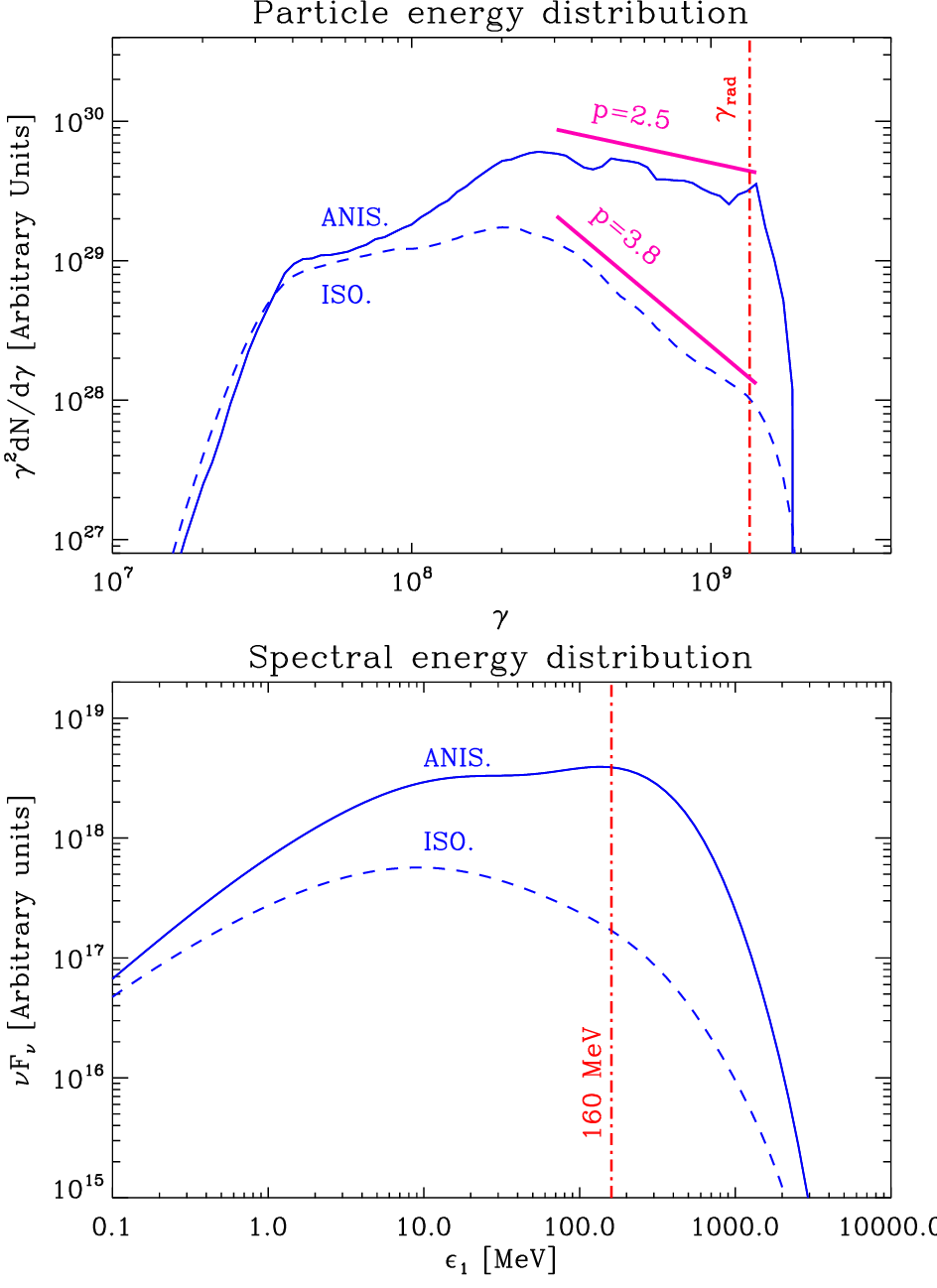}
\caption{Particle ($\gamma^2 d{\rm N}/d\gamma$, top) and spectral ($\nu F_{\nu}$, bottom) energy distributions at $t\omega_1=397$. The solid lines labeled ``ANIS.'' give the distributions as seen by an external observer looking at the direction $\lambda=+70\degr,~\phi=0\degr$ (i.e., in the plane of the reconnection layer, close to the $+x$-direction), within a solid angle $\Delta\Omega/4\pi\approx 3\times 10^{-3}$ (shown by the black box in Fig.~\ref{fig_angular}). The dashed lines are the isotropically averaged distributions labeled ``ISO.'', as shown in Fig.~\ref{fig_spectra_iso} for comparison. In the top panel, the purple segments are power-law fits to the anisotropic and isotropic particle spectra, $d{\rm N}/d\gamma\propto\gamma^{-p}$ between $3\times 10^8<\gamma<1.5\times 10^9$, where $p$ is the best-fit power-law index given above each segment. The red vertical dot-dashed line marks the classical radiation-reaction-limited energy $\gamma_{\rm rad}$ in the top panel, and the corresponding $160~$MeV synchrotron photon energy limit in the bottom panel.}
\label{fig_spectra_anis}
\end{figure}

Fig.~\ref{fig_spectra_iso} presents the energy distributions of all the background particles ($\gamma^2 d{\rm N}/d\gamma$, top panel) and their instantaneous, transparent synchrotron radiation ($\nu F_{\nu}$, bottom panel) at $t\omega_1=0,~110,~220,~397,~$and $662$. The contribution from the drifting particles is not shown here because of their small number compared with the background particles. We assume that the radiation is emitted continuously (valid if $\gamma B\ll B_{\rm QED}$) and tangentially to the particle's orbit (valid if $\gamma\gg 1$). Photons do not interact with the plasma (optically thin approximation), hence there is no need to solve the full radiative transfer equation. Fig.~\ref{fig_spectra_iso} shows the emergence of a high-energy tail of particles, whose maximum energy increases with time during the most active phase of reconnection, $t\omega_1\lesssim 450$ (i.e., during the plasmoid-dominated phase). During the early stage of reconnection, the initial $-2$ power-law extends to higher energies, and cuts off exponentially. The spectrum marginally extends to lower energies. At $t\omega_1\gtrsim 220$, there are some particles that are accelerated above the radiation-reaction limit energy $\gamma_{\rm rad}\approx 1.3\times 10^9$. We find that these extremely energetic particles are accelerated at X-points, surfing multiple times across the reconnection layer, following relativistic Speiser orbits. The details of the Speiser acceleration mechanism are described below in Section~\ref{speiser}. At $t\omega_1=397$, the spectrum above $\gamma=3\times 10^8$ is well described by a steep power-law, i.e., $d{\rm N}/d\gamma\propto\gamma^{-3.8}$, followed by a sharp cut-off beyond $\gamma=\gamma_{\rm rad}$ (Fig.~\ref{fig_spectra_anis}, top panel). We find that about $0.03\%$ of all the particles are above $\gamma_{\rm rad}$, and represent about $0.39\%$ of the total kinetic energy. These particles are responsible for the excess of synchrotron radiation above $160~$MeV (see Fig.~\ref{fig_spectra_iso}), which represents about $4.3\%$ of the total isotropic radiative power. After $t\omega_1=450$, the high-end of the spectrum contracts to lower energies, and very few particles above $\gamma_{\rm rad}$ survive at $t\omega_1\gtrsim 662$. We attribute the disappearance of the most energetic particles to synchrotron cooling. In the final saturated state, most particles are located within the big islands where they cool progressively. There is little acceleration and the magnetic field remains strong, of order $B_0$. If the radiation reaction force is neglected, the high-energy component of the spectrum does not evolve once established (for $t\omega_1\gtrsim 550$, see Fig.~\ref{fig_spectra_iso}), and extends up to $\gamma_{\rm max}\approx 3\times 10^9$. An identical simulation performed without radiative losses with {\tt Vorpal} gives very similar results.

We investigate the angular distribution of the particles' velocities, as a function of their energy. In agreement with our previous study \citep{2012ApJ...754L..33C}, we find a pronounced energy-dependent anisotropy of the particles and their synchrotron emission, increasing with energy. Fig.~\ref{fig_angular} illustrates the strong anisotropy of the expected synchrotron radiation, as a function of the photon energy. Following \citet{2012ApJ...754L..33C}, we use the spherical angles $\phi$ (latitude) and $\lambda$ (longitude) to study the angular distributions. The latitude varies between $-90\degr$ and $+90\degr$, and the longitude varies between $-180\degr$ and $+180\degr$. A radial unit vector has the coordinates $x=\cos\phi\sin\lambda$, $y=\sin\phi$, $z=\cos\phi\cos\lambda$. At $t\omega_1=397$, we find that half of the $>160$~MeV radiative flux is concentrated into less than $4\%$ of the total solid angle $4\pi$. The high-energy beam of radiation is concentrated in the mid-plane ($xz$-plane, $\phi=0\degr$), preferentially towards the $\pm x$-directions ($\phi=0\degr$, $\lambda=\pm 90\degr$). This result can be explained by the deflection of the particles' trajectories from the $\pm z$-directions ($\phi=0\degr,~\lambda=0\degr,~\pm 180\degr$, along which the particles are accelerated by $E_{\rm z}$) to the $\pm x$-directions by the reconnected field. The direction of the beam is changing with time, wiggling around the plane of the reconnection layer during the active phases of reconnection ($t\omega_1\lesssim 450$). The beam broadens and stabilizes along the $z$-direction at later times.

\begin{figure}
\epsscale{1.2}
\plotone{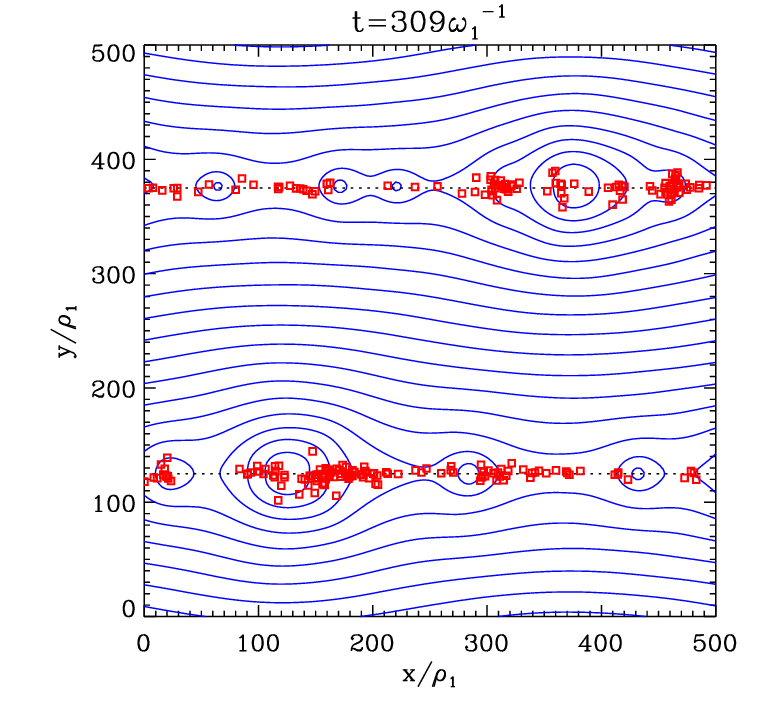}
\caption{Spatial distribution of a random sample of 234 high-energy particles with $\gamma>5\times 10^8$. The red squares show the locations of each these particles in the $xy$-plane at $t\omega_1=309$. The solid lines are the magnetic field lines.}
\label{fig_bunch}
\end{figure}

The strong anisotropy of the emitted radiation leads to an apparent boosting of the flux seen by an observer looking in the direction of the beam (the so-called ``kinetic beaming'', see \citealt{2012ApJ...754L..33C}). The energy distribution of the particles pointing in the direction $\lambda=+70\degr,~\phi=0\degr$ (indicated by the black box in Fig.~\ref{fig_angular}) within the solid angle $\Delta\Omega/4\pi\approx 3\times 10^{-3}$ is substantially harder than the isotropic one. A power-law fit yields $d{\rm N}/d\gamma\propto\gamma^{-2.5}$ above $\gamma=3\times 10^8$ up to $\gamma_{\rm max}\approx\gamma_{\rm rad}$ (Fig.~\ref{fig_spectra_anis}, top panel). In this case, the particles with Lorentz factors above $\gamma_{\rm rad}$ account for about $0.5\%$ of the particles and $8.5\%$ of the energy. Similarly, the apparent spectral energy distribution is dominated by the high-energy radiation, peaking at around $100$~MeV with about $20\%$ of the radiative power above $160~$MeV. The spatial distribution of high-energy particles is strongly inhomogeneous (see also \citealt{2012ApJ...754L..33C}). Fig.~\ref{fig_bunch} shows the spatial distribution of a sample of high-energy particles with $\gamma>5\times 10^8$, at $t\omega_1=309$. The energetic particles are clustered into compact bunches within the reconnection layer and magnetic islands. We note that the high-energy particles are preferentially located at the periphery of the big islands or inside small, newly formed islands. The particles near the centers of big islands are not very energetic because they are no longer accelerated and have cooled radiatively over the time their host island has grown to a large size.

The pronounced inhomogeneity and anisotropy of the extremely energetic particles above $\gamma_{\rm rad}$, and the associated radiation above $160~$MeV, are key elements in explaining the Crab gamma-ray flares (see Section~\ref{crab}).

\subsection{Relativistic Speiser orbits}\label{speiser}

\begin{figure*}
\epsscale{1.0}
\plotone{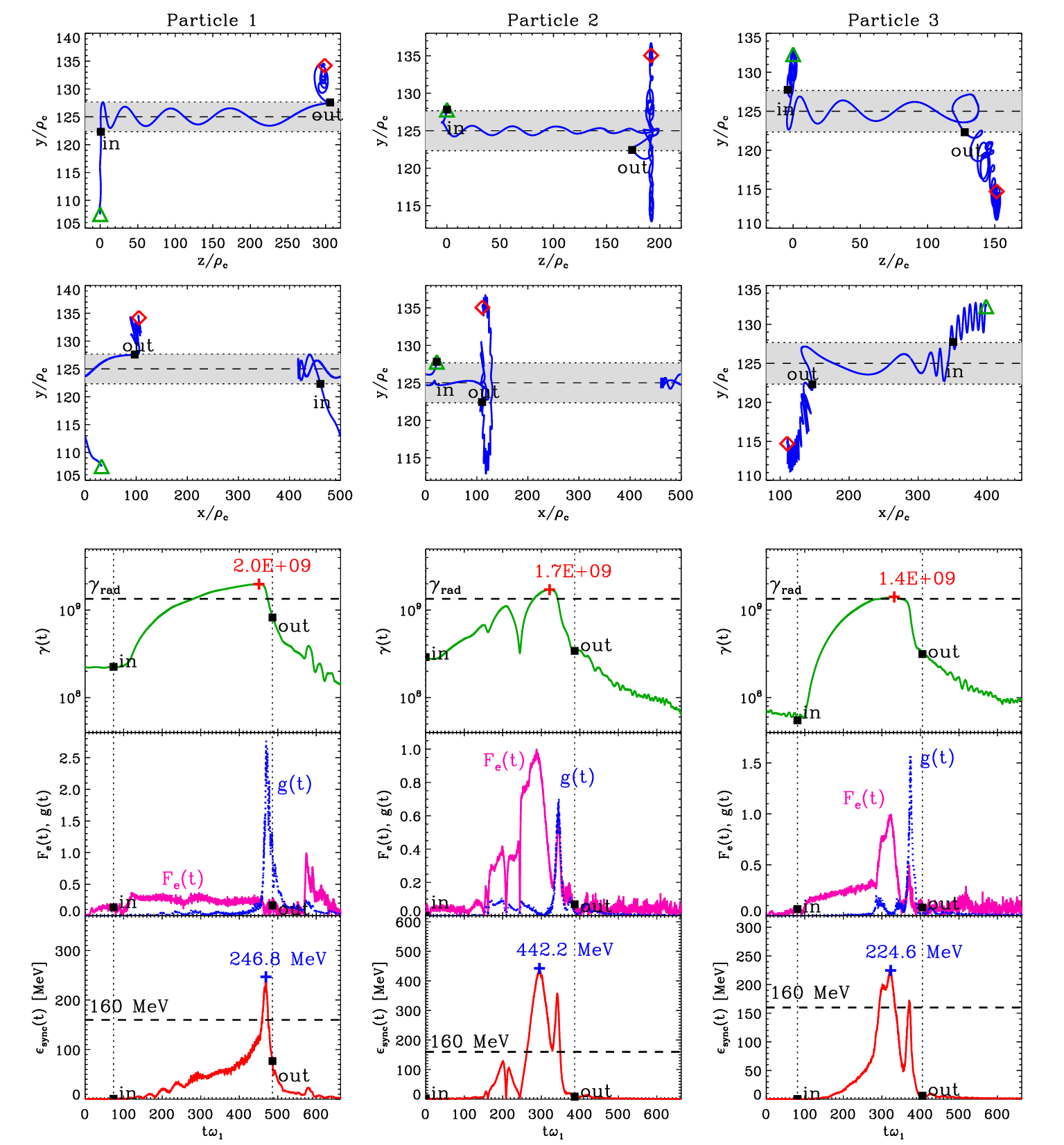}
\caption{Top panels: Three typical trajectories (projected onto the $yz$- and $xy$-planes: note the compressed $y$-axis) of high-energy particles accelerated above the radiation reaction limit, $\gamma_{\rm rad}$, inside the reconnection layer. The gray bands indicate the initial thickness of the layer. The green triangle marks the position of the particle at $t=0$, while the red diamond marks the position of the particle at the end of the simulation at $t\omega_1=662$. Bottom panels: Time evolution along the particle trajectory of (from top to bottom) the Lorentz factor, the strength of the electric force $F_{\rm e}(t)$ (purple solid line) and of the radiation reaction force $g(t)$ (blue dotted line), and the critical synchrotron photon energy $\epsilon_{\rm sync}(t)$. The black squares indicate two reference points along the particle orbit where the particle moves ``in'' and ``out'' of the layer. The crosses mark the maximum values of $\gamma$ and $\epsilon_{\rm sync}$ reached by the 
particle; these values are also printed nearby.}
\label{fig_orbit}
\end{figure*}

In this section, we examine in detail the acceleration mechanism of the most energetic particles with $\gamma>\gamma_{\rm rad}$. We follow the trajectories of $20,000$ particles, picked randomly and uniformly throughout the box at $t=0$. In this sample, there are about a dozen particles reaching a maximum energy above $\gamma_{\rm rad}$. Fig.~\ref{fig_orbit} shows three typical particle trajectories projected in the $yz$-plane and in the $xy$-plane, as well as the time history along the particle trajectory of the particle Lorentz factor $\gamma$, the relative strength of the electric force and the radiation reaction force, and the synchrotron critical energy $\epsilon_{\rm sync}=3heB_{\perp}\gamma^2/4\pi mc$.

These extremely energetic particles systematically follow the same simple time evolution, which can be decomposed into four main phases:

\begin{itemize}

\item{{\bf 1. Drifting:} The particle is initially located upstream and is well magnetized where ideal MHD holds ($E<B_{\perp}$). The particle moves together with the field lines towards the layer, gains little energy and does not radiate much. This phase ends at the reference point ``in'' in Fig.~\ref{fig_orbit} when the particle gets inside the layer.}

\item{{\bf 2. Linear acceleration:} This phase starts when the particle reaches the reconnection layer, where it is no longer magnetized. The strong electric field $E_{\rm z}$ accelerates the particle almost linearly along the $z$-direction. In addition, the magnetic field $B_{\rm x}$ reversing across the layer confines the particle towards the layer mid-plane. The particle follows a relativistic Speiser orbit \citep{1965JGR....70.4219S,2004PhRvL..92r1101K,2007A&A...472..219C,2011ApJ...737L..40U}, whose meandering width $y_{\rm m}$ (maximum distance of the particle from the neutral sheet) decreases with time as the particle energy increases. The particle is then confined closer and closer to the layer mid-plane, where the perpendicular magnetic field strength and hence the radiation reaction force decrease, while the electric force keeps on accelerating the particle. During this process, the particle's trajectory is also significantly deflected towards the $\pm x$-directions by the reconnected field $B_{\rm y}$. This phase begins at the reference point ``in'' and ends when the particle reaches its maximum energy, marked by the red cross in Fig.~\ref{fig_orbit}.}

\item{{\bf 3. Ejection and emission:} While it remains deep inside the layer, the particle is accelerated above the radiation reaction limit $\gamma_{\rm rad}$. The particle is then ejected from the layer, in most cases when it encounters the final big magnetic island where it feels a sharp increase of the radiation reaction force. The big magnetic island acts effectively as a ``beam dump''. The particle loses most of its energy in a fraction of a Larmor cycle and emits synchrotron photons above $160~$MeV. This phase happens just before and after the particle exits the layer (around the reference point ``out'' in Fig.~\ref{fig_orbit}).}

\item{{\bf 4. Cooling:} The particle is back in a region where ideal MHD conditions apply and cools progressively. The particle does not experience any significant acceleration once the saturated state of reconnection is reached.}

\end{itemize}

\begin{figure}
\epsscale{1.1}
\plotone{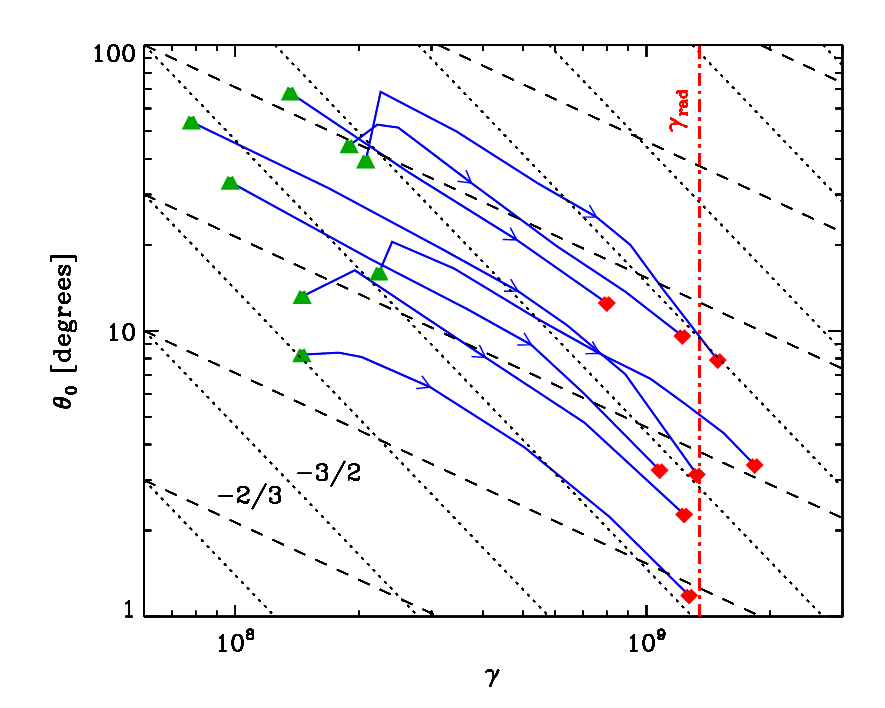}
\caption{Evolution of the particle's mid-plane crossing angle $\theta_0$ with the particle's Lorentz factor $\gamma$, for a representative sample of $8$ high-energy particles accelerated via the Speiser mechanism. The green triangle and the red diamond mark the first and the last crossing of the particle through the layer mid-plane. The particles shown here undergo between $4$ and $9$ crossings before they are kicked outside the layer. The arrow along each particle's path indicates the direction of increasing time. The power-laws of index $-2/3$ (dashed lines) and $-3/2$ (dotted lines) are analytical solutions of relativistic Speiser orbits found by \citet{2011ApJ...737L..40U}. The vertical dot-dashed line shows the radiation reaction limit Lorentz factor $\gamma_{\rm rad}$ (Eq.~\ref{grad}).}
\label{fig_speiser}
\end{figure}

This simple picture highlights the distinction between the acceleration zone (inside the layer) and the $>160~$MeV synchrotron radiating zone (upstream, or inside magnetic islands). The acceleration zone is of order the system size $\sim L_{\rm x}$, while the radiating zone is a fraction of the Larmor radius $\ll\gamma_{\rm rad}mc^2/eB_0$. The acceleration and focusing mechanisms described above in phase 2 agree surprisingly well with our previous test-particle simulations \citep{2011ApJ...737L..40U,2012ApJ...746..148C}, despite the simplistic assumptions on the fields used in these studies (prescribed and static). \citet{2011ApJ...737L..40U} predicted a relationship between $\gamma$ and the angle $\theta_0$ between the particle's velocity vector and the layer mid-plane defined at each crossing, in two extreme regimes. If the particle's meandering width $y_{\rm m}$ is much greater than the layer thickness $\delta$, and if radiative losses are negligible (i.e., during the first Speiser cycles), then $|\theta_0|\propto \gamma^{-2/3}$. In contrast, if the particle is deep inside the layer and reaches the local radiation reaction limit energy $\gamma'_{\rm rad}$ (defined with the perpendicular field at the location of the particle $B_{\perp}<B_0$, so that $\gamma'_{\rm rad}>\gamma_{\rm rad}$) within each cycle, then $|\theta_0|\propto \gamma^{-3/2}$. Fig.~\ref{fig_speiser} shows the tracks followed by a representative sample of $8$ high-energy particles in the $\theta_0$-$\gamma$ plane (which are not necessarily accelerated above $\gamma_{\rm rad}$). The mid-plane crossing angle is given by $\theta_0=\pi/2-\arccos(v_{\rm y}/\sqrt{\mathbf{v}\cdot\mathbf{v}})$, where $\mathbf{v}$ is the three-velocity vector of the particle. The agreement with the analytical expectations is very good: the particles remain between these two power-laws, tending to a $-2/3$ index at low energies and to a $-3/2$ index at the highest energies. This is a robust and clean feature of the most energetic particles accelerated and focused through the Speiser mechanism.

\subsection{Variability pattern of the $>$100~MeV emission}\label{var}

\begin{figure}
\epsscale{1.1}
\plotone{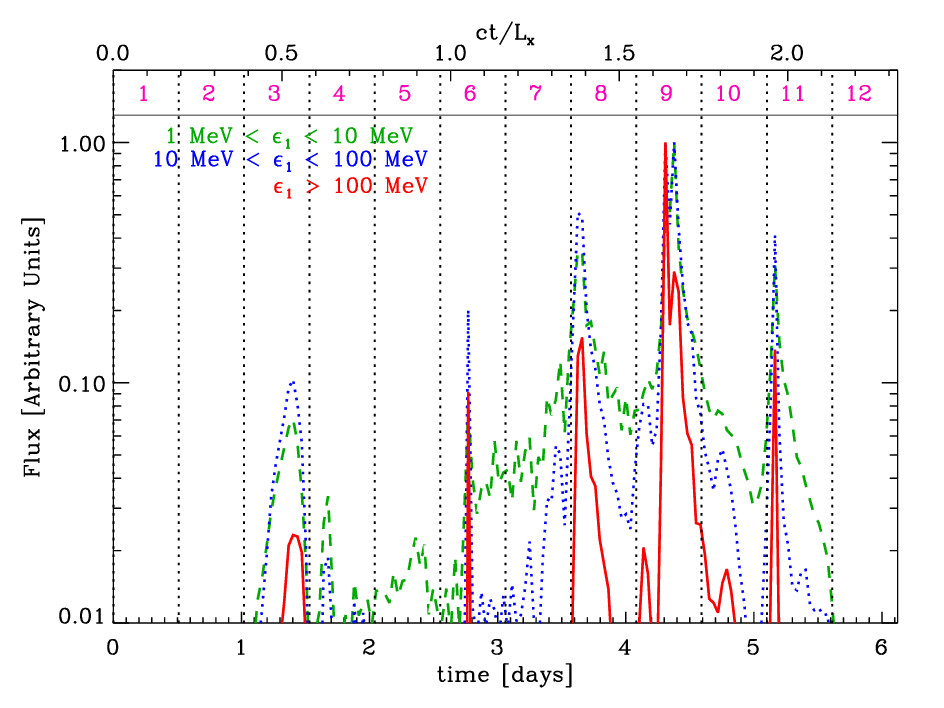}
\caption{Normalized synchrotron flux emitted by the positrons as a function of time (given in days, bottom axis, and in light crossing time of the system, $ct/L_{\rm x}$, top axis) in three photon energy bands: $1$~MeV$<\epsilon_1<10~$MeV (green dashed line), $10$~MeV$<\epsilon_1<100~$MeV (blue dotted line), and $\epsilon_1>100~$MeV (red solid line). The radiation received by the observer is going along the $+x$-direction ($\phi=0\degr,~\lambda=+90\degr$) throughout the simulation within a solid angle $\Delta\Omega\approx 0.03$~Sr. The radiation comes from the bottom layer only. The vertical dotted lines delimit the 12 time periods of equal duration, used to study spectral variability above $100~$MeV in Fig.~\ref{fig_corr}.}
\label{fig_light}
\end{figure}

\begin{figure}
\epsscale{1.1}
\plotone{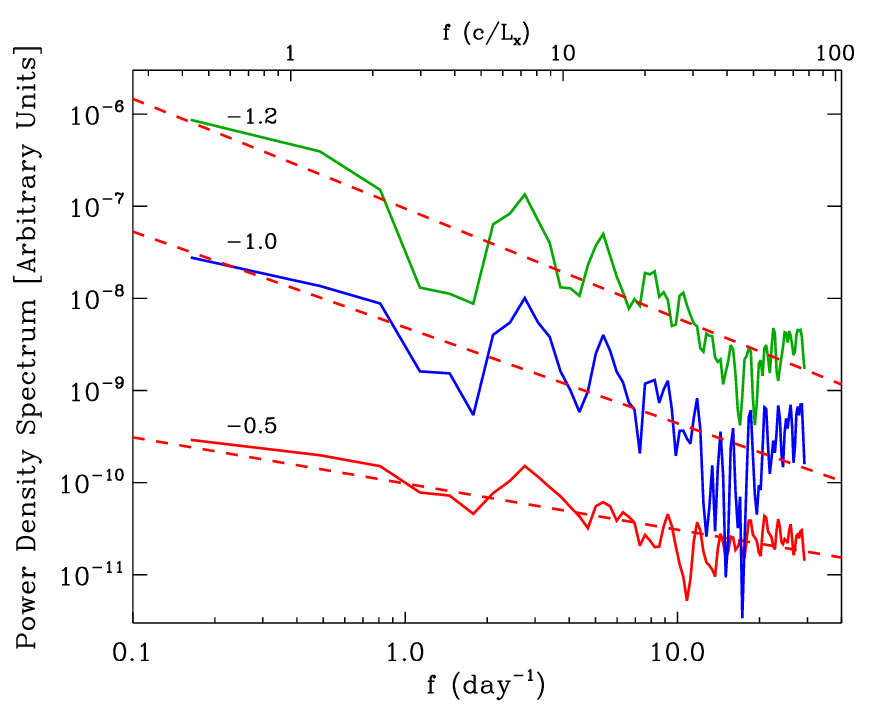}
\caption{Power-density-spectrum of the observed light curve in the three photon energy bands defined in Fig.~\ref{fig_light}: $1$~MeV$<\epsilon_1<10~$MeV (green lines, top), $10$~MeV$<\epsilon_1<100~$MeV (blue lines, middle), and $\epsilon_1>100~$MeV (red lines, bottom). The solid lines give the PDS of the light curve shown in Fig.~\ref{fig_light}. The frequency $f=1/t$ is in day$^{-1}$ (and in units of $c/L_{\rm x}$, top axis), ranging from the inverse of the total duration of the light curve, i.e. $\approx 1/6$ day$^{-1}$, to the inverse of the time resolution of the light curve $\Delta t_{\rm lc}\approx 30~$day$^{-1}$. The red dashed lines are best-fit power-laws of the power-density-spectra, with indexes shown above each line.}
\label{fig_pds}
\end{figure}

In this section, we investigate the time-dependent radiation escaping in the $+x$-direction where most of the high-energy radiation is expected (Fig.~\ref{fig_angular}). Fig.~\ref{fig_light} presents the expected synchrotron flux integrated above $100$~MeV as a function of time, taking into account the time delay due to the light crossing time through the box. In the case of radiation into the $+x$-direction, the propagation time is given by $t_{\rm propag}=(L_{\rm x}-x_{\rm e})/c$, where $x_{\rm e}$ is the location of the emitting electron/positron. In agreement with \citet{2012ApJ...754L..33C}, the high-energy radiation is highly variable on timescales much shorter than the light crossing time of the layer ($\lesssim 0.1 L_{\rm x}/c$, or $\lesssim 6$~hours). The light curve is composed of multiple intense spikes that are nearly symmetric in time. This result is a direct consequence of the strong focusing of the energetic particles accelerated through the Speiser mechanism. The beam of energetic particles wiggles around the reconnection layer and crosses the line of sight several times. The bunching of the high-energy particles into compact blobs within the layer and within the magnetic islands also contributes to the multiple, powerful sub-flares in the light curve \citep{2012ApJ...754L..33C}. This dramatic variability disappears if, instead of considering one particular direction, the emission is averaged over all directions. Fig.~\ref{fig_light} also shows the energy dependence of the light curve. The amplitude of the spikes increases with the energy of the radiation considered, because of the increasing emission anisotropy (Fig.~\ref{fig_angular}). Fig.~\ref{fig_pds} presents the resulting power-density-spectrum (PDS) of the light curve (given by the squared modulus of the Fast-Fourier-Transform), in the three energy bands defined in Fig.~\ref{fig_light}. The observed PDS above $100$~MeV is well-fit by a hard power-law of index $\approx-0.5$. At lower energies, the best-fit indexes are $\approx-1.0$ in the $10~$MeV$<\epsilon_1<100~$MeV band, and $\approx-1.2$ in the $1~$MeV$<\epsilon_1<10~$MeV band. As expected, the PDS slope hardens with increasing photon energy, indicating that the highest energy radiation is also the most rapidly variable.

\begin{figure}
\epsscale{1.1}
\plotone{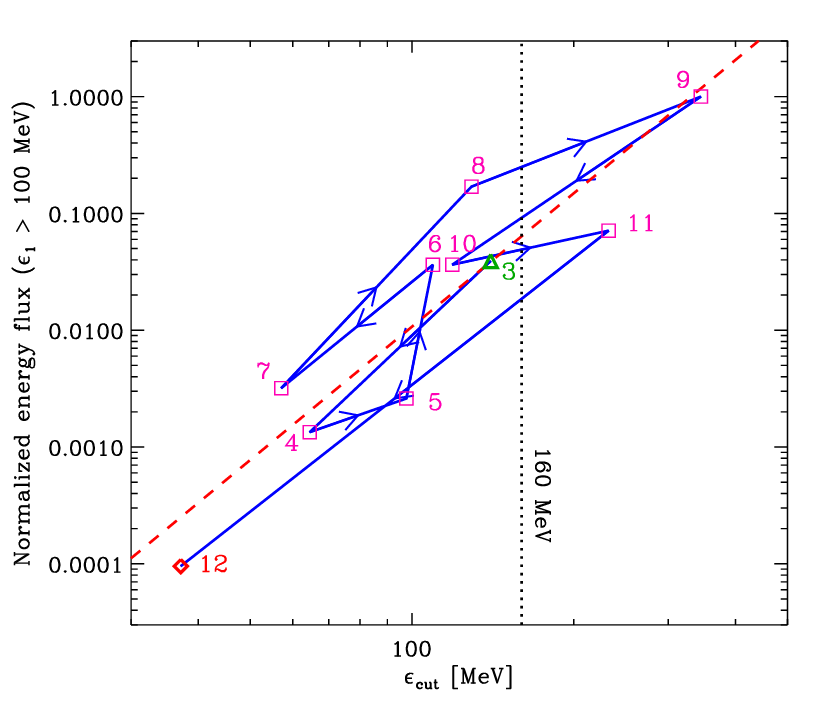}
\caption{Energy flux of the synchrotron radiation above $100$~MeV ($\nu F_{\nu}(\epsilon_1>100~$MeV)) as function of the spectral cut-off energy, $\epsilon_{\rm cut}$, found using the analytical fit $\nu F_{\nu}=K_{\nu}\epsilon_1^{\alpha}\exp\left(-\epsilon_1/\epsilon_{\rm cut}\right)$. Each point corresponds to the observed spectrum averaged over the time periods from $1$ to $12$ defined in the light curve in Fig.~\ref{fig_light}. The periods ``1'' and ``2'' do not appear in this plot because there is no high-energy flux at these times. The arrow shows the path followed over time by the synchrotron spectrum. There is a clear correlation between the energy flux and the cut-off energy. A power-law fit gives $\nu F_{\nu}(\epsilon_1>100~$MeV)$\propto\epsilon_{\rm cut}^{+3.8\pm 0.6}$, overplotted here as a red dashed line. The dotted vertical line marks $\epsilon_{\rm cut}=160~$MeV.}
\label{fig_corr}
\end{figure}

The received spectrum is also highly time variable. We decompose the light curve into $12$ blocks of duration $12~$hours each (see Fig.~\ref{fig_light}). Within each period of time, we compute the time-averaged synchrotron spectrum received by the observer. We fit the high-energy component above $100~$MeV only with a power-law times an exponential cut-off, $\nu F_{\nu}=K_{\epsilon}\epsilon_1^{\alpha}\exp\left(-\epsilon_1/\epsilon_{\rm cut}\right)$, where the free parameters are the normalization constant $K_{\epsilon}$, the spectral index $\alpha$ and the cut-off energy $\epsilon_{\rm cut}$. This analysis reveals a strong correlation between the cut-off energy and the total flux above $100$~MeV (see Fig.~\ref{fig_corr}). A power-law fit gives $\nu F_{\nu}(\epsilon_1>100~$MeV)$\equiv\int_{100~{\rm MeV}}^{+\infty}F_{\nu}d\epsilon_1\propto\epsilon_{\rm cut}^{+3.8\pm 0.6}$. To test the robustness of this correlation, we performed a series of $10$ identical simulations, where only the initial positions and velocities of particles vary. We find a strong and positive flux/cut-off energy correlation in all the simulations, using $50$ time bins to refine our analysis. The power-law index varies from about ${\rm corr}\approx+2$ to ${\rm corr}\approx+4$, with a mean index ${\rm corr}\approx +3$ (see Table~\ref{tab_index}). We note that the slope depends on the low-energy cut-off (100~MeV). However, because the PDS varies significantly from one simulation to another, we are only able to see a hint of a reproducible pattern. On average, the PDS are best fitted by power laws with the following indices: $\approx-1.4$ in the $1~$MeV$<\epsilon_1<10~$MeV band, $\approx-1.3$ in the $10~$MeV$<\epsilon_1<100~$MeV band, and $\approx-1.0$ above $100$~MeV. The deviations from the mean indices within the sample are large, of order $\pm 0.5$.

\begin{table}[htp]
\caption{Correlation between the energy flux and the high-energy photon spectral cut-off energy, for a statistical sample of $10$ identical simulations where only the initial positions and velocities of particles vary. The correlation coefficient is the best-fit power-law index as shown in Fig.~\ref{fig_corr}, i.e., $\nu F_{\nu}(\epsilon_1>100~$MeV)$\propto\epsilon_{\rm cut}^{\rm corr}$, using $50$ bins in time.}
\label{tab_index}
\centering
\begin{tabular}{|c|c|}
\hline
Simulation & Best-fit index $\rm corr$ \\
\hline
1 & $3.14\pm 0.45$\\
2 & $2.57\pm 0.44$\\
3 & $3.26\pm 0.41$\\
4 & $2.90\pm 0.50$\\
5 & $1.96\pm 0.38$\\
6 & $4.07\pm 0.43$\\
7 & $3.33\pm 0.52$\\
8 & $2.89\pm 0.39$\\
9 & $3.05\pm 0.37$\\
10 & $3.44\pm 0.56$\\
\hline
\end{tabular}
\end{table}

\subsection{Effect of the guide field}\label{guide}

\begin{figure*}
\epsscale{1.0}
\plotone{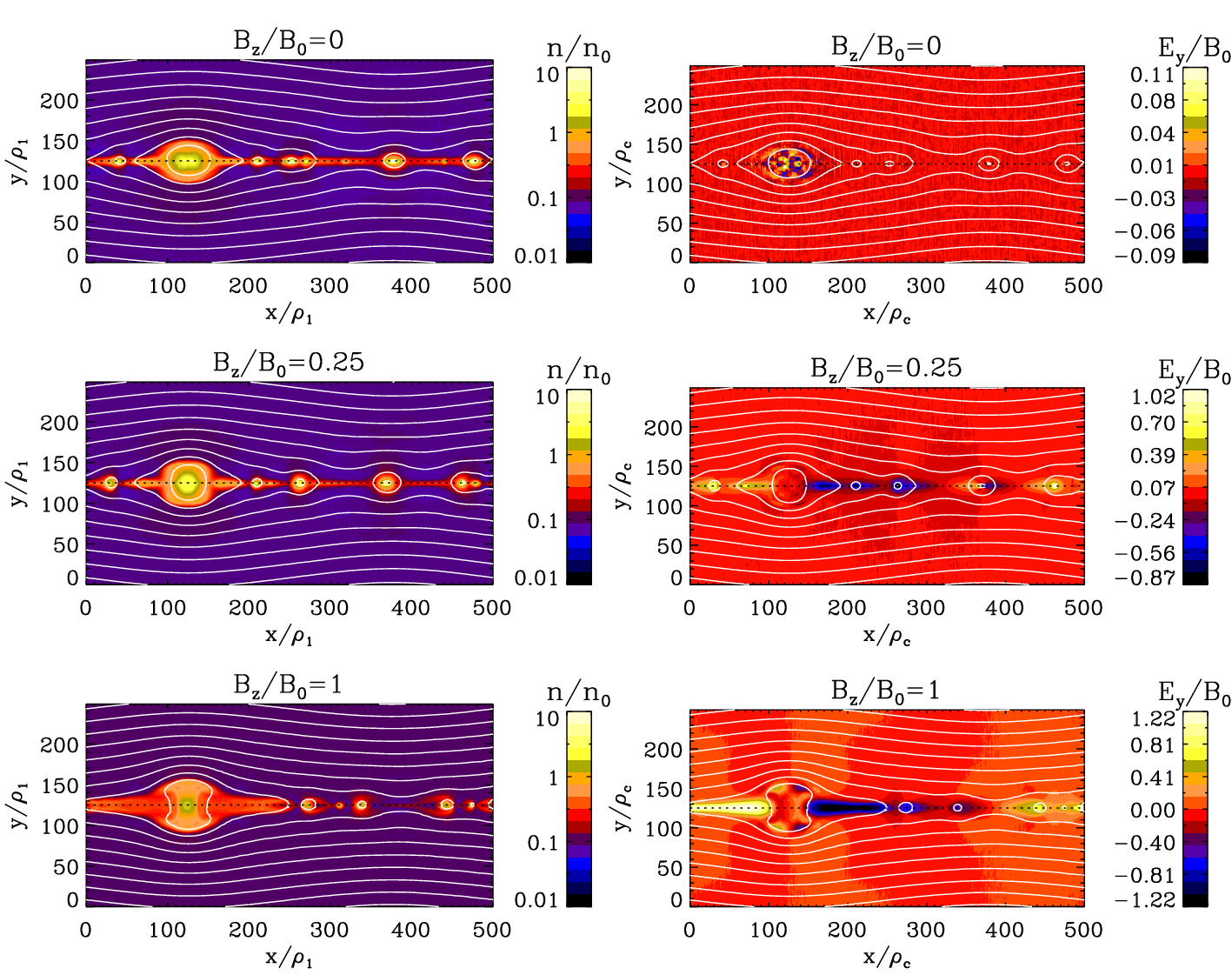}
\caption{Spatial distribution of the plasma density (left panels), and electric field strength $E_{\rm y}$ (right panels) for three different values of the guide field strength, $B_{\rm z}/B_0=0$ (top), $0.25$ (middle), and $1$ (bottom), at $t\omega_1=221$. The figure shows the bottom layer only. The plasma density is normalized to the initial drifting particle density inside the layer, $n_0$, and the electric field is normalized to $B_0$. White solid lines are magnetic field lines in the $xy$-plane.}
\label{fig_density_guide}
\end{figure*}

All the simulations presented above assumed a zero magnetic guide field component $B_{\rm z}=0$. In the general case, however, reconnection is not purely anti-parallel and there is always a finite guide field. In this section, we examine the effect of a moderate and uniform guide magnetic field on particle acceleration and radiation, by performing a set of $4$ simulations with $B_{\rm z}/B_0=0.1,~0.25$, $0.5$, and $1$, with the physical parameters given in Table~\ref{tab_params}. The general time evolution of the reconnection dynamics described in Section~\ref{evol} is still valid with a guide field. However, we find that the morphology of the particle distribution inside the plasmoids is more complex. In the presence of a finite guide field, two streams of particles form symmetrically on both sides of the layer, in the opposite direction to the island motion (Fig.~\ref{fig_density_guide}, left panels). Each arm is produced by the deflection of the particles by the guide field in the $\pm y$-directions, caused by the Lorentz force, and is composed of electrons or positrons only. As a result, the guide field creates a separation of charges within each island across the layer, which in return induces a strong electric field $E_{\rm y}$ of order $B_0$ across the layer (see Fig.~\ref{fig_density_guide}, right panels). We also note the induction of an $E_{\rm x}$ electric field of order $0.1B_0$, reversing across the layer, due to the motion of the incoming plasma towards the layer at a velocity of order $\beta_{\rm rec}c$ in the $\pm y$-directions. We also find that the strong energy-dependent anisotropy of the particles and radiation is preserved with a guide field, although the structure of the beam is more complex (high-energy particles occasionally point towards high latitudes $|\phi|>0\degr$).

\begin{figure}
\epsscale{1.1}
\plotone{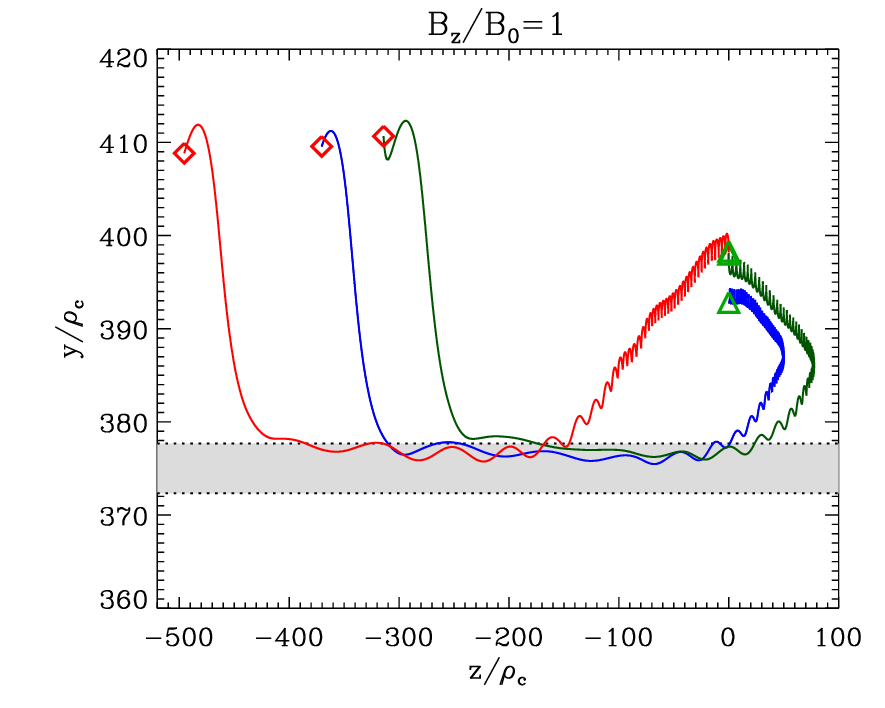}
\caption{Three typical high-energy particle trajectories, projected onto the $yz$-plane, in the presence of a $B_{\rm z}=B_0$ uniform guide field. The initial ($t\omega_1=0$) and final ($t\omega_1=662$) positions of the particles are marked by the green triangles and the red diamonds respectively. The gray band depicts the initial upper-layer thickness $2\delta$.}
\label{fig_guide_orbit}
\end{figure}

\begin{figure}
\epsscale{1.1}
\plotone{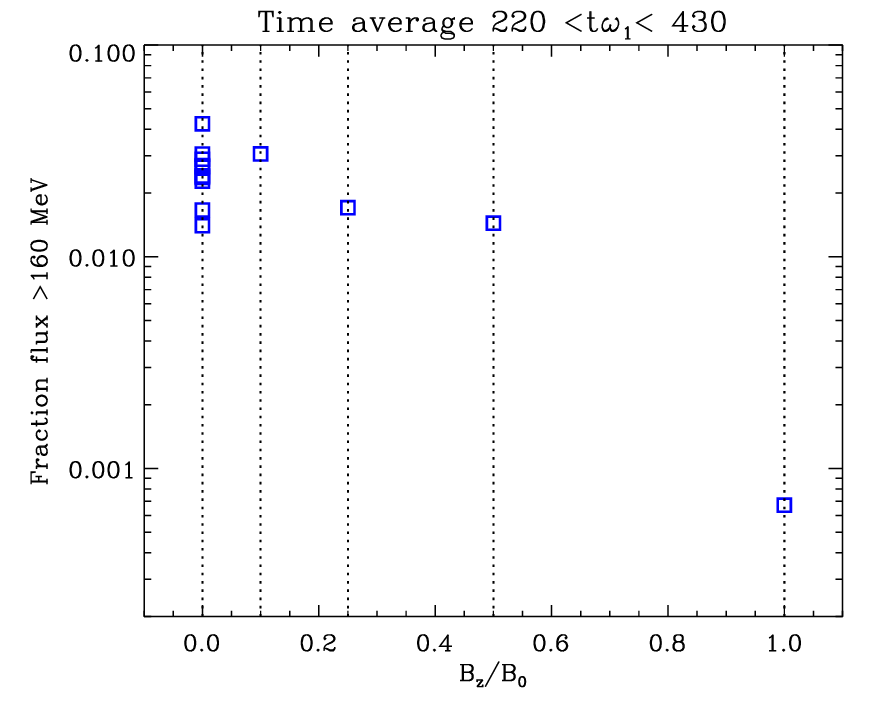}
\caption{Fraction of the total isotropic radiative flux emitted above $160~$MeV, as a function of the guide field strength. The flux is time-averaged over the most active phase of reconnection, i.e. $221<t\omega_1<442$. The blue squares represent the result from each simulation, where $B_{\rm z}/B_0=0,~0.1,~0.25,~0.5,~$and $1$. This figure also shows the variations of the flux above $160~$MeV within the statistical sample of $10$ identical simulations (see Section~\ref{var}), for $B_{\rm z}=0$.}
\label{fig_guide_he}
\end{figure}

The deflection of the particles away from the layer does not limit significantly the efficiency of particle acceleration through the Speiser mechanism. The particle energy distribution extends up to $\gamma\approx 2\times 10^9$. Analysis of a sample of particles tracked throughout the simulation shows that the energetic particles are inevitably deflected away from the layer, but only after several crossings of the layer. Fig.~\ref{fig_guide_orbit} shows three typical high-energy particle trajectories, projected in the $yz$-plane. The particles are progressively carried outside the layer by the guide field, and have time to cool over a Larmor timescale. The radiation reaction force gradually overcomes the electric force because the pitch angle of the particle to the magnetic field line (and hence the perpendicular magnetic field) increases progressively as it exits the layer. In contrast, the ejection of the energetic particles into strong $B_{\perp}$ in the weak guide field case is abrupt: the radiation reaction force jumps rapidly over a short period of time and exceeds the electric force, leading to a rapid (sub-Larmor) cooling and the emission of $>160~$MeV synchrotron photons (Fig.~\ref{fig_orbit}). Hence, the presence of a strong guide field ($B_{\rm z}\gtrsim B_0$) effectively inhibits the emission of $>160~$MeV synchrotron radiation. Fig.~\ref{fig_guide_he} shows the fraction of the total radiative flux emitted above $160~$MeV, as a function of the guide field strength, from $B_{\rm z}=0$ to $B_{\rm z}=B_0$. For $B_{\rm z}=B_0$, the $>160~$MeV flux is about $50$ times smaller than the zero-guide field case. We note that the $>160~$MeV flux varies by a factor $2$--$3$ within our statistical sample of $10$ simulations with no guide field introduced in Section~\ref{var}, as illustrated in Fig.~\ref{fig_guide_he}.

A moderate guide field may be beneficial for particle acceleration in three-dimensional pair plasma reconnection. It tends to suppress the development of the drift kink instability that occurs only in 3D, which may quench non-thermal particle acceleration (\citealt{2008ApJ...677..530Z}, see also the discussion in \citealt{2011PhPl...18e2105L,2011ApJ...741...39S, 2012arXiv1208.0849K}). A large 3D simulation such as those presented here in 2D is currently beyond our computational resources. We leave this issue to a future study.

\section{Solution to Crab gamma-ray flares?}\label{crab}

In this section, we discuss the implications of our findings in the context of the Crab gamma-ray flares. Below, we review the observational features of the flares, and show how reconnection can naturally explain them:

\begin{itemize}

\item{{\bf Synchrotron radiation $\mathbf{>160}$~MeV:} All flares show synchrotron radiation well above the $160~$MeV burnoff limit, up to $375~$MeV during the April 2011 super-flare \citep{2012ApJ...749...26B}. We showed in Section~\ref{speiser} that particles can be accelerated well above the classical radiation reaction limit, via the relativistic Speiser mechanism deep inside the layer, and radiate synchrotron radiation up to $\gtrsim 400$~MeV. The most energetic particles go through a substantial fraction of the electric potential drop available in the simulation, i.e. $\mathcal{E}_{\rm max}\sim e\beta_{\rm rec}B_0 L_{\rm x}\approx 3~$PeV, with $\beta_{\rm rec}=E_{\rm z}/B_0\approx 0.3$, $B_0=5~$mG, and $L_{\rm x}=2.7$~light-days.}

\item{{\bf Hard spectrum}: The April 2011 flare unambiguously shows an extra synchrotron component on top of the quiescent emission. The spectrum averaged over the flare is hard, such that $F_{\nu}\propto\nu^{-0.27\pm 0.12}$ \citep{2012ApJ...749...26B}, i.e., consistent with an emitting population of pairs distributed with a power-law of index $p\approx 1.6$. The energy of the distribution is then dominated by the highest energy particles. We find that a hard spectrum (although not as hard as observed) is naturally expected in our simulations during the brightest periods of high-energy emission (periods 8-9-10 in Fig.~\ref{fig_light}), when $F_{\nu}\propto\nu^{-0.4},~\nu^{-0.5}$ above $100~$MeV. A hard spectrum could explain why there is no detectable counterpart at lower energies (X-rays, optical, IR, radio).}

\item{{\bf Ultra-rapid time variability:} The first short intra-flare variability was detected in the 4-day-long September 2010 flare, over a $12$-hours timescale \citep{2011A&A...527L...4B}. During the 9-day-long April 2011 flare, there are significant variations of the flux over periods $\lesssim 8$~hours \citep{2012ApJ...749...26B}. Hence, there are even smaller structures in the flaring region emitting each spike of high-energy radiation. Along the lines of \citet{2012ApJ...754L..33C}, we find in Section~\ref{var} that super-fast time variability of the observed high-energy flux is expected, due to the strong inhomogeneity and anisotropy of the most energetic particles responsible for the $>100$~MeV emission. We see a flare on Earth only when the beam of high-energy radiation crosses our line of sight. The symmetry of the sub-flare time profile is also consistent with observations. We attribute the overall duration of the Crab flares (from a few days to a couple of weeks) to the reconnection timescale, i.e., of order the light crossing time of the reconnecting region, $6~$days in our simulations (Fig.~\ref{fig_light}). We associate the intra-flare time variability (hours to days) to the light crossing time of magnetic islands generated by the tearing instability, $\lesssim 6$~hours in the simulations (see also \citealt{2013MNRAS.431..355G} for a similar interpretation in the context of super-fast TeV flares in blazars). Our study also reveals that the power-density spectrum of the $>100~$MeV light curve is expected to be a hard power-law of index $\approx -1\pm 0.5$ (Fig.~\ref{fig_pds}). This is consistent with observations that show a power-law index $\approx -1$ above the noise floor \citep{2012ApJ...749...26B}.}

\item{{\bf Flux/energy correlation:} One of the most remarkable features of the April 2011 flare is the discovery of a clear correlation between the $>100$~MeV flux and the cut-off energy, such that $\nu F_{\nu}(>100~$MeV)$\propto \epsilon_{\rm cut}^{+3.42\pm 0.86}$ \citep{2012ApJ...749...26B}. This result is interpreted by \citet{2012ApJ...749...26B} as the signature of the rapid variations of relativistic Doppler-boosted emission \citep{1985ApJ...295..358L}. We discovered in Section~\ref{var} that reconnection can reproduce such a correlation as well, with a power-law index ${\rm corr}\approx+3$, and mimic the effect of Doppler beaming. Relativistic bulk outflows with Lorentz factor of order $\Gamma\sim\sqrt{\sigma}=4$ are expected in relativistic Petschek reconnection \citep{2005MNRAS.358..113L}, and lead to a Doppler amplification of the observed emission \citep{2009MNRAS.395L..29G,2010MNRAS.402.1649G,2011MNRAS.413..333N, 2012MNRAS.426.1374C,2013MNRAS.431..355G}. However, a Petschek-like reconnection configuration is traditionally associated with either the Hall effect or a highly-localized anomalous resistivity due to current-driven plasma microinstabilities, both of which are absent in our 2D pair plasma simulations\footnote{See, however, \citet{2007PhPl...14e6503B} who pointed out that the divergence of the electron/positron pressure tensor can effectively play the role of anomalous resistivity, facilitating fast reconnection in collisionless pair plasmas.}. Hence, the scaling $\Gamma\sim\sqrt{\sigma}$ may not be valid in the plasmoid-dominated reconnection regime explored by the simulations. Our results suggest that the reconnection outflows are not strongly relativistic, i.e., $\Gamma\sim 1$, as expected in relativistic Sweet-Parker reconnection \citep{2005MNRAS.358..113L}. Here, we attribute the beaming of the radiation to the strong energy-dependent anisotropy of the particles only (Section~\ref{spec}, see also \citealt{2012ApJ...754L..33C}). The beaming pattern obtained in the simulations is not consistent with a relativistic Doppler effect that beams all the emission by the same factor, regardless of the particle energy. This is a possible observational test of the kinetic beaming.}

\end{itemize}

\section{Summary}\label{ccl}

The main result of this paper is the first discovery of particle acceleration above the classical synchrotron radiation reaction limit, and the associated emission of $>160~$MeV synchrotron radiation, in numerical simulations of collisionless pair plasma reconnection. For this purpose, we developed a new, parallel, relativistic PIC code, called {\tt Zeltron}, that includes the effects of the radiation reaction force on particle dynamics self-consistently. We confirm in every detail the expectations of earlier (semi-)analytical works \citep{2004PhRvL..92r1101K,2007A&A...472..219C,2011ApJ...737L..40U, 2012ApJ...746..148C}: the most energetic particles are almost linearly accelerated and confined deep inside the reconnection layer, where radiative losses are small while the accelerating electric field is strong. If the particle stays long enough within the layer, it can be accelerated above the radiation reaction limit defined in the upstream plasma. Then, once the particle eventually escapes the layer, it radiates $>160~$MeV synchrotron radiation within a fraction of a Larmor gyration. The acceleration mechanism that emerges from this study is remarkably simple and robust.

Although this work addresses a fundamental question in particle acceleration, it is mainly motivated by the mystery of the Crab Nebula gamma-ray flares. We have shown in this study that all the puzzling aspects of the flares are consistent with a reconnection event in the nebula. In addition to the $>160$~MeV synchrotron radiation, our simulations can explain the ultra-rapid time variability of the observed high-energy radiation. The emission originates mostly from extremely anisotropic and compact bunches of energetic particles. An external observer sees a flare when the beam of energetic particles crosses the line of sight. We found that the power-spectrum of the light curve is well-fit by a power-law, which shows hints of hardening with the energy band of the radiation. In addition, we discovered that there is a strong positive correlation between the emitted radiative flux and the high-energy spectral cut-off, mimicking the effect of a relativistic Doppler boost. This correlation is also consistent with the observations of the Crab flares. A strong guide field, i.e. $\gtrsim B_0$, deflects particles out of the reconnection layer and tends to suppress the emission $>160~$MeV synchrotron radiation. Our results support the magnetic reconnection scenario at the origin of the Crab flares.

\acknowledgements We thank Rolf Buehler, Dimitrios Giannios, Geoffroy Lesur, Krzysztof Nalewajko, Anatoly Spitkovsky, and the referee for helpful discussions. This research was supported by an allocation of advanced computing resources provided by the National Science Foundation, by NSF grant PHY-0903851, NSF grant AST-0907872, DoE grant DE-SC0008409, and NASA Astrophysics Theory Program grant NNX09AG02G. Numerical simulations were performed on the local CIPS computer cluster Verus, on Kraken at the National Institute for Computational Sciences (\url{www.nics.tennessee.edu/}). This work also utilized the Janus supercomputer, which is supported by the National Science Foundation (award number CNS-0821794), the University of Colorado Boulder, the University of Colorado Denver, and the National Center for Atmospheric Research. The Janus supercomputer is operated by the University of Colorado Boulder.

\end{document}